\documentclass[12pt]{article}

\usepackage{newtxtext,newtxmath}
\usepackage{xcolor}
\usepackage{soul}
\usepackage{comment}
\usepackage{graphicx}

\usepackage[letterpaper,margin=1in]{geometry}

\renewenvironment{abstract}
	{\quotation}
	{\endquotation}

\date{}

\makeatletter
\renewcommand{\fnum@figure}{\textbf{Figure \thefigure}}
\renewcommand{\fnum@table}{\textbf{Table \thetable}}
\makeatother

\usepackage{scicite}

\usepackage{url}

\newcommand{\Chen}[1]{\textcolor{blue}{#1}}

\def\scititle{ Attosecond-resolved quantum fluctuations of light and matter
}
\title{\bfseries \boldmath \scititle}

\author{
	Matan~Even~Tzur$^{1*\dagger}$,
	Chen~Mor$^{2\dagger}$,
	Noa~Yaffe$^{2}$,
        Michael~Birk$^{1}$,
        Andrei~Rasputnyi$^{3,4}$,\and
        Omer~Kneller$^{2}$,
        Ido~Nisim$^{1}$, 
        Ido~Kaminer$^{1}$,
        Maria~Chekhova$^{1,3,4}$,
        Michael~Krüger$^{1}$,\and
        Misha Ivanov $^{1,5}$,
        Nirit~Dudovich$^{2}$,
        and Oren~Cohen$^{1,6}$
        \and
	\small$^{1}$Technion - Israel Institute of Technology, Haifa, Israel\and
	\small$^{2}$Department of Physics of Complex Systems, Weizmann Institute of Science, Rehovot, Israel\and
        \small$^{3}$Max Planck Institute for the Science of Light, Erlangen, Germany\and
                \small $^{4}$Friedrich-Alexander Universität Erlangen-Nürnberg
        \and 
        \small$^{5}$Max Born Institute, Berlin, Germany
        \and
        \small$^{6}$ Guangdong Technion-Israel Institute of Technology, Shantou, Guangdong 515063, China
   \\ 
	\small$^\ast$Corresponding author. Email: matan.tzur@mpsd.mpg.de\and
	\small$^\dagger$These authors contributed equally to this work.
}

\newcommand{\gt}{$g^{(2)} $ }
\newcommand{\gtm}{g^{(2)} }
\newcommand{\avg}[1]{\langle {#1} \rangle}

\newcommand{\Var}[1]{\text{Var}({#1})}
\newcommand{\Cov}[1]{\text{Cov}({#1})}
\newcommand{\beq}{\begin{equation}}
\newcommand{\eeq}{\end{equation}}

\begin{document} 
\maketitle

\begin{abstract} \bfseries \boldmath
 Until recently, attosecond optical spectroscopy and quantum optics evolved along non-overlapping directions. In attosecond science, attosecond pulses have been regarded as classical waves, applied to probe electron dynamics on their natural time scale. Here, we transfer fundamental concepts of quantum optics into attosecond physics, enabling control of both the properties of the XUV attosecond pulses and the quantum fluctuations of matter on attosecond time scales. 
 By combining bright squeezed vacuum (BSV) with a strong laser field to drive high-harmonic generation, we transfer the 
 quantum properties of the BSV onto the resulting XUV attosecond pulses. Applying advanced attosecond interferometry, we reconstruct the quantum state of the XUV high harmonics and their associated attosecond pulses with attosecond precision. Finally, we resolve the squeezing of the electron's wavepacket during one of the most fundamental strong-field phenomena -- field induced tunneling. The ability to measure and control quantum correlations in both electrons and XUV attosecond pulses establishes a foundation for attosecond quantum electrodynamics, manipulating the quantum state of electrons and photons with sub-cycle precision.

\end{abstract}

\noindent 

Attosecond spectroscopy has revolutionized our ability to capture fundamental ultrafast phenomena, resolving electron dynamics on their natural time scale \cite{Villeneuve2018}. The foundation and driving force of this field is the generation of XUV attosecond pulses via a strong-field light-matter interacti on, leading to high harmonic generation (HHG) \cite{Ferray1988,Li1989}. Over the past two decades HHG-based spectroscopy has resolved, with unprecedented precision, a range of fundamental quantum phenomena, from field induced tunneling \cite{Pedatzur2015} or multi-electron dynamics in atoms \cite{shiner2011probing},
to charge migration in molecules \cite{ChargeMigration,smirnova2009high} and topological dynamics in solids \cite{TopologicalHHG}. 
These experiments share one common characteristic -- the \textit{classical} properties of attosecond pulses served as a probe to resolve or manipulate the \textit{quantum} dynamics of electronic wavefunctions.  While the measurement and control of the classical properties of attosecond pulses and their associated XUV high harmonics are established \cite{Gariepy2014,Fleischer2014, huang2018polarization}, characterization and control of their quantum properties remains a major challenge.
Building on these advances, the next frontier lies in extending attosecond spectroscopy beyond classical observables to the quantum domain of light itself. Resolving this challenge is a key step towards interfacing attosecond technology and spectroscopy with quantum optics, opening a new field of ultrafast quantum-optical spectroscopy. 

Quantum light \cite{loudon2000quantum}, predominantly produced in the visible to infrared spectral range, plays a critical role in quantum information processing \cite{Braunstein2005}, precision metrology \cite{Aasi2013}, and photonic quantum technologies \cite{PhotonicQuantumTech}. While most quantum light sources are weak, BSV \cite{Chekhova2015} - a microjoule-level squeezed state with nonclassical correlations - stands out as a unique source to study  the interaction between intense quantum light and matter. 
Employing BSV to drive harmonic generation in the UV regime \cite{Rasputnyi2024, Lemieux2024} and multiphoton photoemission \cite{Heimerl2024,pollothdriving} is one route to extend quantum optics into the ultrafast, high-photon-energy regime.
Non-classical optical signatures of harmonics have also been observed during harmonic generation driven by coherent states of light, either in the outgoing infrared driving field \cite{Lewenstein2021} or in the visible harmonics \cite{Theidel2024}.
However, so far,  the quantum state of XUV high harmonics and attosecond pulses, as well as attosecond-scale quantum fluctuations in matter, have remained unexplored.

Recent theoretical studies suggest that applying quantum optical concepts to high harmonic generation opens a door to a new class of quantum phenomena where the quantum properties of light manipulate matter, or are controlled by it, 
on a sub-cycle time scale \cite{Gorlach2020,Lewenstein2021, Gorlach2023, Harrison2023, EvenTzur2023, Tzur2024, Sloan2023, Lange2024, Yi2024, RiveraDean2024,rivera2024non}. It has been proposed that HHG can generate entangled XUV photon pairs \cite{Sloan2023}, massively entangled \cite{Lange2024,Yi2024}, and squeezed light states\cite{Tzur2024,RiveraDean2024}, unveil 
non-adiabatic material excitations \cite{yi2025generation} and many-body correlations in correlated materials \cite{Lange2024} as well as in superradiating ensembles \cite{pizzi2023light}. 
These remarkable opportunities pose critical challenges: Can the quantum state of high harmonics and the associated attosecond pulses be characterized and controlled, in the XUV range with attosecond precision? What is the link between their quantum state and the field driven dynamical electronic wavefunction that has led to their generation?

Here, we demonstrate the generation of an attosecond pulse train in the XUV regime with controllable quantum properties by manipulating electronic wavefunction correlations with sub-cycle precision. We combine intense 800 nm laser light with relatively weak quantum light generated by a 1600 nm BSV source to produce XUV emission with tailored quantum characteristics. These properties are controlled by imprinting the quantum fluctuations of the BSV field onto the electronic wavefunction during light-induced tunneling, with attosecond precision, and subsequently transferred onto the quantum properties of the emitted XUV pulses. To resolve the quantum state of the XUV light, we developed a novel form of attosecond quantum tomography performed in-situ -- during the generation process itself -- where individual atoms generating the XUV emission serve as interferometers \cite{dudovich2006measuring,Pedatzur2015}.  Hence, our scheme enables the first reconstruction of the quantum state of XUV light, as represented by the Wigner or Husimi distribution \cite{schleich2015quantum}. 
Finally, we demonstrate that the quantum properties of the emitted light encode quantum fluctuations in matter at previously inaccessible temporal and energy scales, revealing fluctuations arising during light-induced electron tunneling. Our study transfers quantum optics to new extremes, opening new opportunities for attosecond quantum electrodynamics (QED).

	\begin{figure}[!htb]
		\begin{center}
		
		\centering{\includegraphics*[width=0.9\columnwidth]{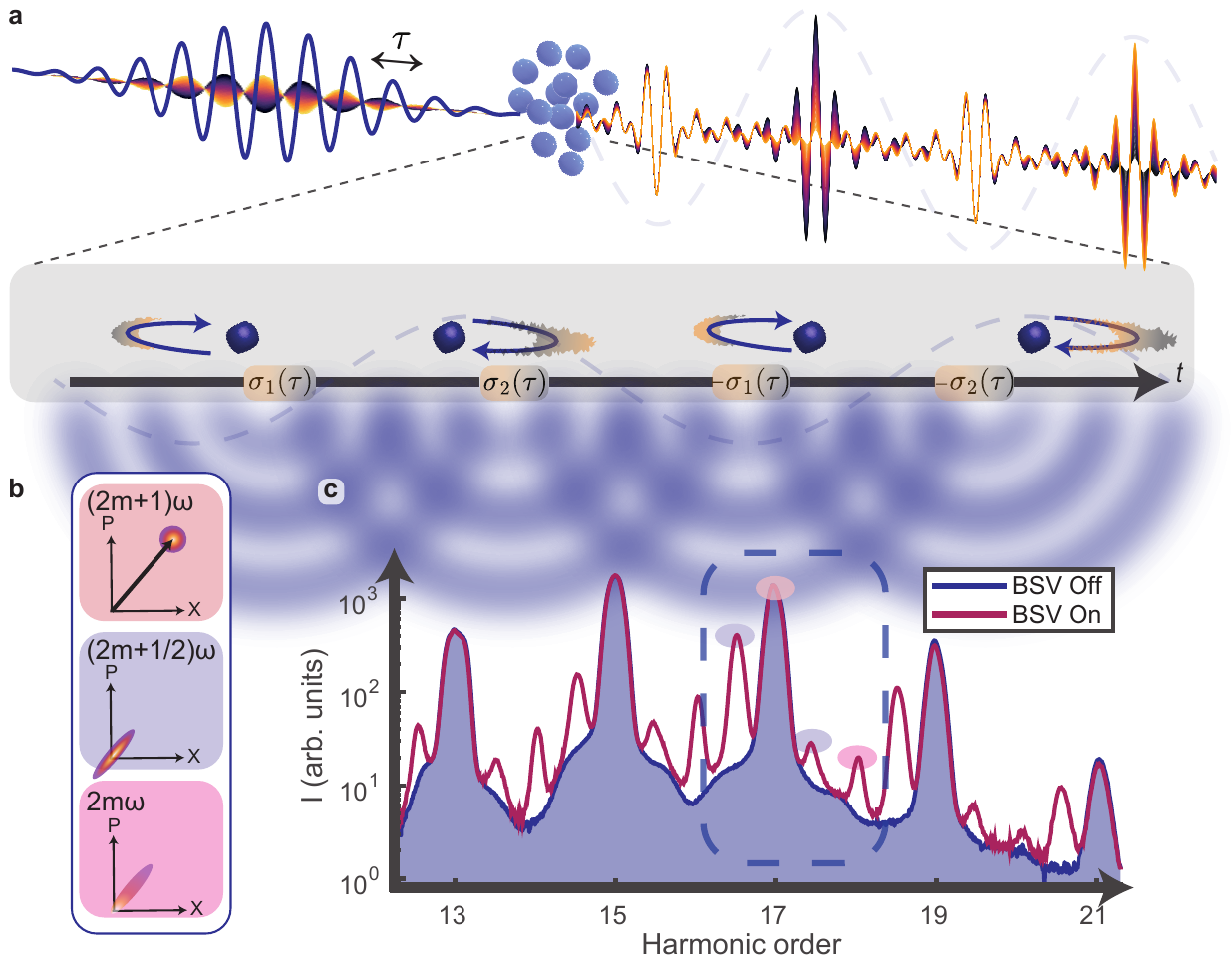}}
		\caption{ {\bf Generation of squeezed attosecond XUV pulses}. {\bf a},  HHG is driven by the combination of a strong coherent field of frequency \(\omega\) and a weak bright squeezed vacuum (BSV) field of frequency \(\omega/2\), producing a train of quantum attosecond pulses. The \(\omega-\omega/2\) geometry acts as a temporal analogue of a four-slit interferometer, where four attosecond bursts are emitted per \(\omega/2\) cycle. Each burst has a fluctuating amplitude and phase described by a complex phase shift \(\sigma_1(\tau)\) and \(\sigma_2(\tau)\), representing the perturbation of the BSV to the quantum action of the electron. Their interference gives rise to a comb of integer and half-integer squeezed harmonics.
        {\bf b}, Schematic illustrations of phase space diagrams for harmonics of different orders. Odd harmonics are approximately in a coherent state, half-integer harmonics exhibit squeezed fluctuations, and even harmonics exhibit phase-space displacement and squeezed fluctuations. {\bf c}, Experimentally resolved HHG spectra generated by a coherent field only (blue) and the two-color field, composed of coherent and BSV sources (red). The perturbative field leads to the appearance of half-integer (blue) and even (pink) harmonics.}
			\label{firstFigure}
		\end{center}
	\end{figure}

\subsection*{Attosecond quantum interferometry}
Controlling and characterizing the quantum properties of attosecond pulses imposes a significant challenge. Attosecond pulses are generated during strong-field light-matter interaction, involving large photon numbers, while quantum light commonly involves low photon numbers. We overcome this challenge by combining a strong coherent beam centered at $800$ nm with a weaker  BSV beam centered at $1600$ nm. The coherent field governs the strong-field dynamics, while the BSV field acts as a perturbation, imprinting its non-classical characteristics onto the correlations and statistical properties of the generated attosecond pulses. 

Our measurement and control scheme leverages the well-established three-step model of gaseous HHG \cite{corkum2007attosecond,schafer1993above}: within each half-cycle of the fundamental 800 nm field ($\omega$) an electron tunnel-ionizes in the strong driving field, propagates in the continuum, and recombines with its parent ion, emitting an XUV attosecond pulse. 
When driven solely by a fundamental $800$ nm field ($\omega$), the symmetry between consecutive half cycles ensures a constructive interference at odd harmonics and destructive interference at even harmonics\cite{BenTal1993}. Introducing a $1600$ nm perturbation with a frequency \(\omega/2\) breaks this symmetry, leading to the generation of half-integer and even harmonic orders. Quantum-mechanically, the properties of the harmonics are governed by the action accumulated along the quantum electron trajectory \cite{lewenstein1994theory,salieres2001feynman}. A weak 1600 nm field perturbs this action \cite{Pedatzur2015,dudovich2006measuring,shafir2012resolving,dahlstrom2011quantum}, imposing a complex phase shift \(\sigma_j =\alpha_j +i \beta_j\) ,
associated with each fundamental half cycle \(j\) \cite{dahlstrom2011quantum}. The real part of the perturbation, \(\alpha_j\), describes the change in the phase accumulated along the quantum electron  trajectory, while its imaginary part, \(\beta_j\), represents the change in the amplitude of this trajectory, dominated by the perturbation of the ionization step. For a sufficiently weak $1600$nm field, these shifts are linear with the perturbing field amplitude. The perturbation exhibits a periodicity spanning four half cycles of the driving 800 nm field, with symmetry that dictates the relationships among the four complex  phases: 
\(\sigma_1; \sigma_2; \sigma_3 = -\sigma_1;\sigma_4 = -\sigma_2\) (Figure \ref{firstFigure}a). This configuration serves as a four-ports in-situ interferometer, realized on a single atom level. The complex phase perturbation modifies the odd harmonics ($2N+1$) and 
generates new frequency components: the half-integer harmonics ($2N\pm{\frac{1}{2}}$) and even harmonics ($2N$), as directly observed in the measured spectra (Figure 1c). The corresponding intensity modulations are (see Supplementary Material):

\begin{equation}
I_q \propto |E_q|^2 \propto
\begin{cases} 
    \left| \cos( \sigma_1) + \cos( \sigma_2) \right|^2, & \text{q=2N+1}, \\
    \left| \cos( \sigma_1) - \cos( \sigma_2) \right|^2, & \text{q=2N}, \\
    \left| i\sin( \sigma_1) + \sin( \sigma_2) \right|^2, & 
    \text{q=2N} + \frac{1}{2}, \\
    \left| i\sin( \sigma_1) - \sin( \sigma_2) \right|^2, & 
    \text{q=2N} - \frac{1}{2}.
\end{cases}
\label{eq:I}
\end{equation}

In contrast to previous works \cite{Pedatzur2015,luu2018observing, worner2010following,he2010interference, dahlstrom2011quantum}, in our experiment $\sigma_1$ and $\sigma_2$ as well as the harmonic intensities are inherently stochastic variables. The physical origin of their fluctuations is the noise of the non-classical BSV, initiated by the fluctuations of the quantum vacuum. In the absence of the BSV perturbation, the odd harmonics are in coherent states \cite{rivera2022strong}, a property that remains approximately valid for a sufficiently weak BSV perturbation. In contrast, since $\sigma_1$ and $\sigma_2$ are approximately linear in the perturbative field,  Eq. 1 shows that half-integer harmonics are also approximately linear in the perturbative field in the regime where \(sin(\sigma_{1,2})\approx\sigma_{1,2}\). 
Hence, they are anticipated to resemble the BSV statistics, exhibiting vanishing displacement in phase space and squeezed fluctuations \cite{Tzur2024} (Figure \ref{eq:I}b). Even harmonics are quadratic with the perturbation and are thus expected to be in more complex states, exhibiting non-Gaussian fluctuations.

Equation \ref{eq:I} maps the quantum noise of the electronic wavefunction (encoded in the perturbations to the electron's action, \(\sigma_{1,2}\)) into measurable statistics of the harmonic intensities, and vice versa. As we shall show below, we utilize this mapping to explore the quantum correlations of the electronic wavefunctions as well as to reconstruct the quantum states of the XUV high harmonics. 
The above equation reflects a fundamental property of the interferometric scheme: although the experimental measurement is macroscopic, the interference mechanism originates at the single-atom level, where the average number of generated photons is far below one. Therefore, the in-situ light characterization is effectively performed in the single-photon regime.

\subsection*{Photon statistics}
Photon statistics characterize light by describing its photon number distribution and correlations, serving as one of the key methods for identifying the non-classical fingerprint of light. We first explore the photon statistics of the HHG radiation, dictated by the interplay between the strong-field interaction and the BSV perturbation. A pioneering study successfully measured the photon statistics and intensity–intensity correlations of individual harmonics in the UV range, offering a static measurement at a fixed two-color delay\cite{Lemieux2024}. Here, we resolve these properties in the XUV spectral range and in the attosecond regime, by directly shaping the sub-cycle structure of the two-color driving field with attosecond precision. As expected, the BSV source is characterized by super-Poissonian photon statistics, exhibiting a long tail extending towards high photon numbers (Supplementary Information). We resolve the photon statistics of the high harmonics by conducting single-shot measurements of the XUV spectrum, obtaining the complete photon statistics at a fixed two-color delay (Supplementary Information).
  
	\begin{figure}[!htb]
		\begin{center}
		\centering{\includegraphics*[width=1\columnwidth]{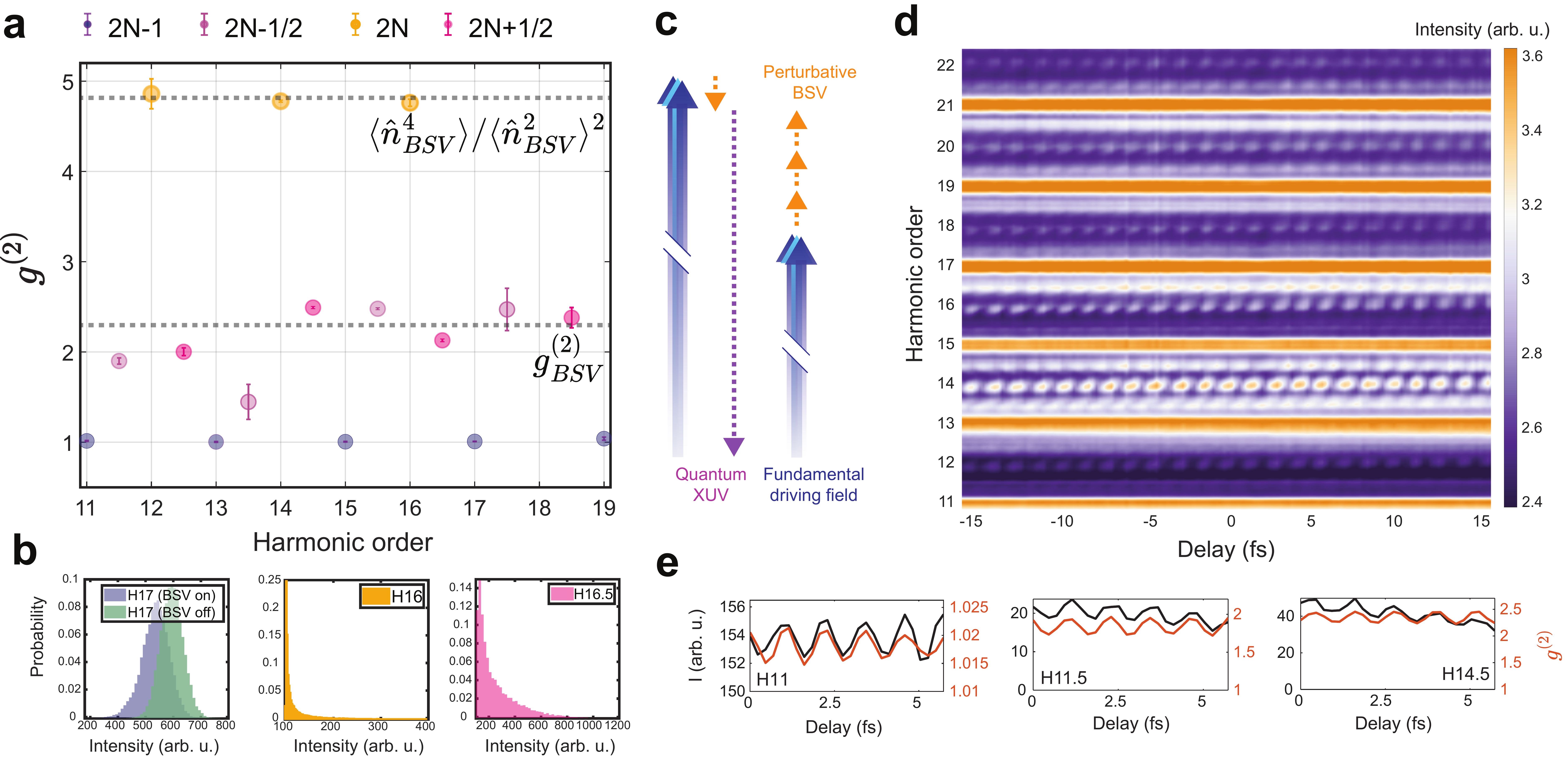}}
		\caption{{\bf Photon statistics of the XUV harmonics for a fixed and scanned two-color delay.} \textbf{a}, Second order correlation function (\gt) as a function of harmonic order. Odd harmonics exhibit Gaussian statistics, with $\gtm\approx 1$. Half-integer and even harmonics exhibit a long-tailed distribution, with \gt values clustering around 2.3 and 4.8, respectively. These values correspond to \gt and the photon number kurtosis $\langle\hat{n}^4\rangle/\langle\hat{n}^2\rangle^2$ for the input BSV, as indicated by dashed gray lines in \textbf{a} (Supplementary Information). \textbf{b} Intensity distributions of three types of harmonics: $2N+1$ (17), 2N (16), $2N + \frac{1}{2}$ (16.5).  {\bf c}, One and three BSV-photon (orange dashed arrows) processes interfere to generate half-integer harmonics. {\bf d}, Mean value of different optical frequencies as a function of the two-color delay. {\bf e}, \gt and mean oscillations of selected harmonics (11,11.5, 14.5 ,15.5)  as a function of the two-color delay (Supplementary Information). The oscillations of \gt are in phase with the mean value oscillations. Notably, very weak, yet clearly resolved, \gt oscillations are observed in harmonic 11.}  
			\label{StatsFig}
		\end{center}
	\end{figure}

As described in Eq. \ref{eq:I}, the photon statistics of the harmonics group into four distinct families -- odd, even, and half-integer harmonics (Figs \ref{StatsFig}a-b). Figure \ref{StatsFig}b presents the photon statistics of harmonic 17, which follows a Gaussian distribution in the absence of the BSV source, as expected when HHG is driven by a  classical coherent state field \cite{Gorlach2020,Lewenstein2021}. Introducing the BSV modifies this distribution, reducing the mean intensity and producing a broader, asymmetric profile, with the longer tail extending towards lower photon numbers. This behavior reflects the anti-correlation between odd harmonic intensities and the perturbation amplitude, as predicted by Eq. \ref{eq:I}. 
Figure \ref{StatsFig}b also presents the photon statistics of exemplary half-integer and even harmonics. These harmonic families exhibit photon statistics similar to those of the input BSV, with a long tail extending towards higher photon numbers.  
To quantify XUV intensity fluctuations, we calculate the second-order  coherence, defined for a large photon number $\avg{\hat{n}}$ as $g^{(2)}= \langle\hat{n}^2-\hat{n}\rangle/\langle\hat{n}\rangle^2 \approx \langle\hat{n}^2\rangle/\langle\hat{n}\rangle^2 $  observing super-bunching ($\gtm > 2 $) in both half-integer and even harmonics (Figure  \ref{StatsFig}a). Half-integer harmonics cluster around $\gtm \approx2.3$, matching the input BSV value (Supplementary Information), as they are mainly produced by a single-BSV-photon process \cite{MultiphotonMasha}. Even harmonics exhibit $g^{(2)}\approx 4.8$, matching the value of the photon number kurtosis $\langle\hat{n}^4\rangle/\langle\hat{n}^2\rangle^2$ for the input BSV, consistent with their generation by a two-BSV-photon process \cite{MultiphotonMasha}. 

So far, we explored the photon statistics of the harmonics for a fixed two-color delay,
revealing, for the first time, the mapping of squeezed vacuum photon statistics into the XUV harmonics. By measuring the single shot statistical distribution for each two-color delay, we can track the sub-cycle dynamics that underlie this process. We observe periodic modulations of the mean intensity with a periodicity of $400$nm, corresponding to half the fundamental optical cycle (Figure  \ref{StatsFig}d-e).
Statistical analysis shows that as we scan the two-color delay, \gt is modulated as well, in phase with the modulations of the mean value of the harmonics (Figure \ref{StatsFig}e). What is the origin of these oscillations? 

These oscillations originate from the modulations of sub-cycle trajectories induced by the two-color field. While the BSV carries a strongly fluctuating amplitude, it maintains a well defined phase. Therefore, such a perturbation leads to amplitude fluctuations (i.e., the perturbation is complex) in each of the four slits of our temporal interferometer, which are mutually correlated and phase-locked with respect to the coherent driving field.  Consequently, the interference between the slits depends on the two-color delay, yielding the observed phase-locked mean intensity and \gt modulations. A complementary view emerges in the frequency domain, where the perturbative nature of the BSV interaction allows us to describe these time domain modulations in terms of photonic pathways. Expanding Eq.  \ref{eq:I} as a power series in the complex phases $\sigma_{1,2}$ describes the interaction as a sum over perturbative BSV photonic processes. Thus, we can view the mechanism as a frequency-domain interference of photonic pathways whose complex amplitudes depend on powers of $\sigma_{1,2}$. 
For example, half-integer harmonics oscillations originate from a combination of one-BSV-photon and three-BSV-photon processes  (Figure \ref{StatsFig}.c), as seen by expanding $\sin(\sigma_{1,2})\approx\sigma_{1,2}-\sigma_{1,2}^3/6$ in Eq. \ref{eq:I}. These pathways, $\sigma$ and $\sigma^3$, acquire different phases as we scan the two-color delay, leading to the modulations of the interference signal. Since the one and three-photon processes have related yet distinct statistics,  \gt   of their interference depends on their relative phase.

In addition, odd harmonic oscillations (Figure  \ref{StatsFig}e), originate from the interference between a coherent state (zero-BSV-photons) and a four-BSV-photon process, as indicated by the lowest-order $\phi$-dependent contribution scaling as $\sigma^4$ (supplementary information). Altogether, these results demonstrate how scanning the two-color delay uncovers interference among multiple photonic pathways, enabling precise control and characterization of the statistical properties of the emitted attosecond pulses.


\subsection*{\NoCaseChange{Reconstruction of tunneling dynamics driven by squeezed light}}
\begin{figure}[!htb]
		\begin{center}
		
		\centering{\includegraphics*[width=0.65\columnwidth]{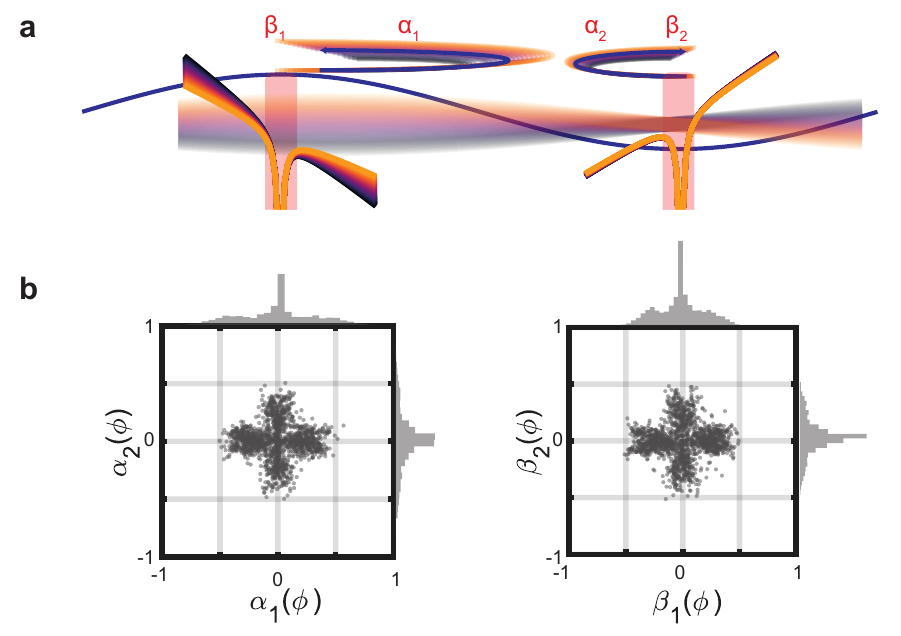}}
		\caption{ {\bf Reconstruction of sub-cycle quantum fluctuations of tunneling in atoms driven by squeezed light} {\bf a}, The coherent field (blue) induces pairs of trajectories, labeled 1 and 2, every half cycle of the fundamental field. The correlation between $\beta_1$ and $\beta_2$ reflects a correlation between two instantaneous tunneling events: when the first exhibits excess noise (a broader $\beta_1$ spread), the subsequent event, half a cycle later, becomes quieter (a narrower $\beta_2$ spread).
        {\bf b}, The correlation between $\alpha_1$ and $\alpha_2$ reflects a correlation between two successive trajectories when the first exhibits excess noise (a broader $\alpha_1$ spread), the subsequent event, half a cycle later, becomes quieter (a narrower $\alpha_2$ spread).
        } 
			\label{ElectroCorrelations}
		\end{center}
	\end{figure}

Our ability to reveal the quantum fluctuations of the harmonics and to resolve their evolution with attosecond precision provides deep insight into the underlying dynamics, allowing us -- for the first time -- to uncover the quantum fluctuations imprinted in the process of quantum tunneling. Quantum tunneling is one of the most fundamental phenomena distinguishing quantum from classical mechanics \cite{landau2013quantum}. It plays a central role across physics, chemistry, and technology, with applications ranging from semiconductor devices to superconducting circuits \cite{devoret2013superconducting,blais2021circuit}. While electron tunneling is inherently quantum -- where an electronic wavepacket propagates through a classically forbidden region -- the tunneling barrier itself has traditionally been regarded as purely classical \cite{keldysh2024ionization,perelomov1966ionization,ammosov1986tunnel,torlina2012time}.
In this work, we explore electronic tunneling through a barrier modulated by quantum noise, implemented by the photon statistics of our BSV source.
According to Eq. \ref{eq:I}, the statistics of the recombining HHG electrons over four consecutive half-cycles -- represented by the distributions of  $\sigma_{1,2}$ -- are mapped onto the joint intensity statistics of four adjacent harmonics (for instance, 14.5, 15, 15.5, and 16). Thus, by inverting Eq. \ref{eq:I}, we can extract two complex electron phases $\sigma_{1,2}\equiv\alpha_{1,2}+i\beta_{1,2}$ on a shot-by-shot basis from the intensities of four adjacent harmonics (supplementary information). Such analysis reconstructs the joint statistics of $\sigma_{1,2}$, for each two-color delay.        
Figure \ref{ElectroCorrelations}a schematically describes the origin of the real and imaginary components of the sub-cycle perturbations to the action, highlighting the stochastic nature of the perturbation. The imaginary phase, $\beta_j$, corresponds to a modification of the tunneling, as the BSV modifies the ionization barrier near the peak of the coherent driving field, at half-cycle \(j\). The real term, $\alpha_j$, represents the additional phase accumulated by the electronic wavefunction between ionization and recombination, due to its interaction with the BSV field. 

Performing single-shot measurements of the harmonic spectrum allows us to reconstruct the statistical distribution of these complex phases as well as their correlations along successive half-cycles.
For example, the $\beta_1,\beta_2$ correlations, evident in Figure \ref{ElectroCorrelations}b reveals that when one tunneling event exhibits excess noise (resulting in a broad $\beta_1$ spread), the subsequent event, occurring half a cycle later, experiences reduced noise (leading to a narrower $\beta_2$ spread). This pattern reflects the temporal interval between excess and reduced noise in the input BSV, controlled by the delay between the BSV and the coherent field. Retrieving its temporal dependence constitutes the first demonstration of a tunneling barrier with tunable quantum properties. 

\subsection*{\NoCaseChange{Quantum state tomography}}

We reconstruct the quantum state of the emitted harmonic field by mapping, via Eq. (1), the joint statistics of the complex electronic phases $\sigma_{1,2}$ to the Husimi Q representation - the probability distribution to measure a coherent state \(\alpha=x+ip\) with quadrature amplitudes \(x\) and \(p\): 

\begin{equation}
Q(\alpha) = \frac{1}{\pi} \langle \alpha | \hat{\rho} | \alpha \rangle, 
\qquad \alpha = x + ip
\label{eq:Husimi}
\end{equation}

Here, $\hat{\rho}$ is the density operator and $|\alpha\rangle$ a coherent state with quadrature amplitudes $x$ and $p$. Experimentally, we obtain $Q(\alpha)$ directly: for each laser shot, the complex harmonic amplitude $E_q \propto x + i p$ is computed from Eq.
~\eqref{eq:I} using the extracted phases $\sigma_{1,2}$. A two-dimensional histogram of the resulting $\alpha = (x,p)$ values over many shots then yields the Husimi distribution $Q(\alpha)$.

\begin{figure}[!htb]
		\begin{center}
		
		\centering{\includegraphics*[width=0.95\columnwidth]{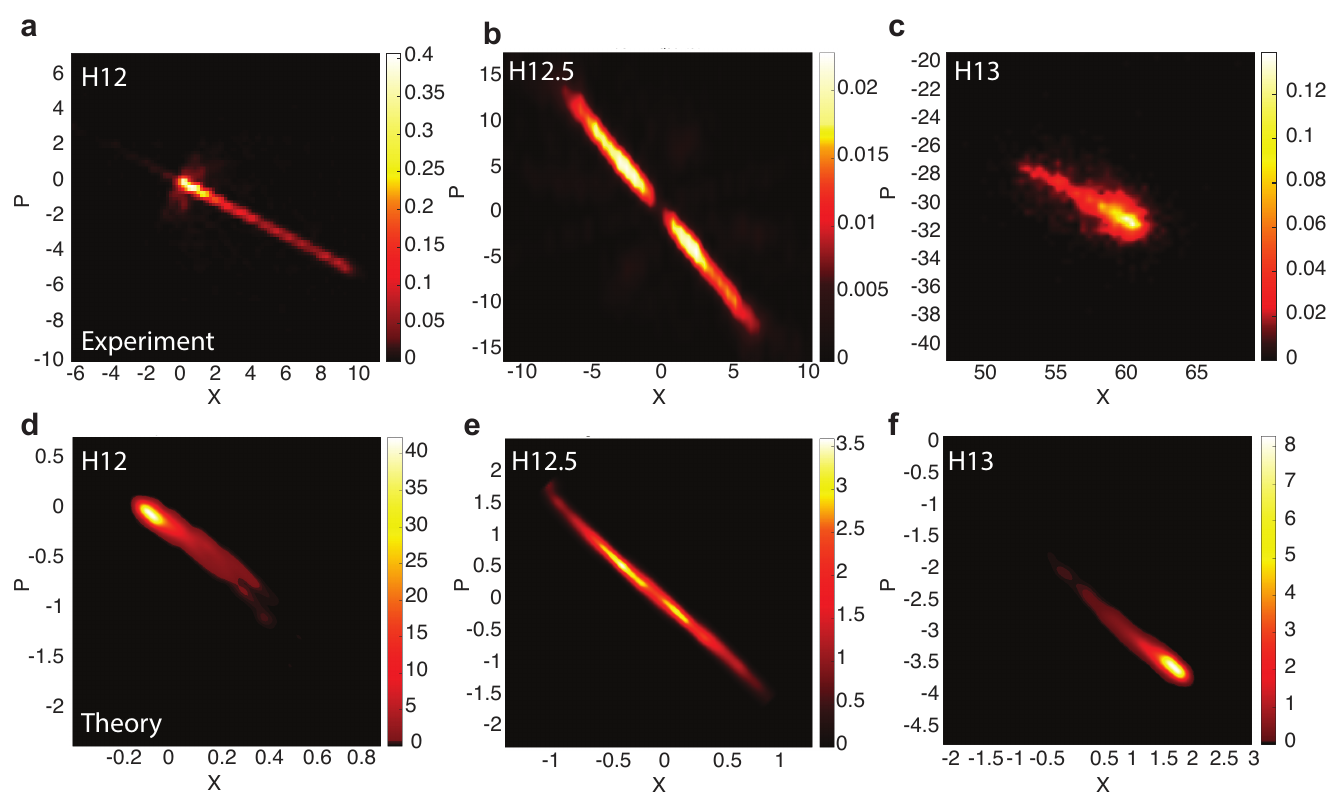}}
		\caption{ {\bf Quantum state tomography of high-harmonic emission.} (a–c) Experimentally reconstructed Husimi distributions for harmonics at $\lambda=66$nm (harmonic 12), 64nm (harmonic 12.5), and 61.5nm (harmonic 13). (d–f) Corresponding theoretical calculations (supplementary information). {\bf a,d} Even harmonic orders (12) appears as a squeezed-like state elongated along one quadrature. {\bf b,e} Half-integer harmonic orders (12.5) exhibits a non-Gaussian, two-lobed structure with a central hole, reflecting the nonlinear mapping of the Gaussian BSV fluctuations into the harmonic quadratures. {\bf c,f} Odd harmonic orders (13) show a dominant coherent-like lobe with a trailing depletion toward the origin of phase space, consistent with energy redistribution into neighboring harmonics half-integer and even harmonics. The close correspondence between experiment and theory highlights the robustness of our interferometric quantum-state tomography scheme.}
			\label{fig:Husimi}
		\end{center}
	\end{figure}

Representative results are shown in Figure \ref{fig:Husimi} for harmonic orders 12 (even), 12.5 (half-integer), and 13 (odd), alongside theoretical predictions.
Interestingly, in contrast with the input coherent and BSV fields, the measured and calculated Husimi distributions of the half-integer harmonic (H12.5, panels b,e) are clearly non-Gaussian, displaying a zero-displaced elliptical envelope with two lobes and a central hole. The elliptical shape is a hallmark of squeezed vacuum, reflecting the imbalance between squeezed and anti-squeezed quadratures. The emergence of two peaks and the central hole indicates a departure from Gaussianity, reminiscent of a cat-like superposition. 
This behavior follows from Eq.\eqref{eq:I}: the half-integer harmonics depend nonlinearly on the BSV field via $\sin(\sigma_{1,2})$. This nonlinear dependence of the harmonic response distorts the squeezed fluctuation distribution of the BSV field into the observed non-Gaussian, squeezed-cat-like state.
The Husimi distribution of the even harmonic (H12, panels a,d) exhibits a state reminiscent of the input BSV. However, rather than forming a symmetric distribution around the origin of phase space, it is displaced and stretched predominantly along a single direction, while its most probable value remains near zero. The pronounced elongation along one quadrature axis reflects the quadratic dependence of even harmonics on the perturbing BSV field. In contrast, the Husimi distribution of the odd harmonic (H13, panels c,f) displays a dominant lobe corresponding to the BSV-unperturbed coherent-state emission, accompanied by a tail extending toward the origin of phase space. This tail signifies a reduction of the unperturbed coherent-state emission as energy is transferred into the half-integer and even harmonics. 
Altogether, these observations highlight the rich variety of quantum states accessible across this unique XUV HHG spectrum.

To the best of our knowledge, this is the first experimental demonstration of quantum-state tomography in the XUV spectral range (as represented by the Husimi distribution, see Supplementary Information for the Wigner-function reconstruction).
This reconstruction highlights the application of self-referenced internal interferometry which serves as a viable substitute for conventional homodyne tomography at attosecond time scales. We note that this scheme is complementary to quantum state tomography of attosecond electrons \cite{priebe2017attosecond,bourassin2020quantifying,koll2022experimental,laurell2025measuring}. This observation holds great potential for the development of XUV quantum sources. By spectrally filtering quantum HHG radiation, one could realize a tunable quantum source, paving the way for a broad range of future applications. 

\subsection*{\NoCaseChange{Time-Domain Reconstruction}}
\begin{figure}[!htb]
		\begin{center}
		
		\centering{\includegraphics*[width=0.65\columnwidth]{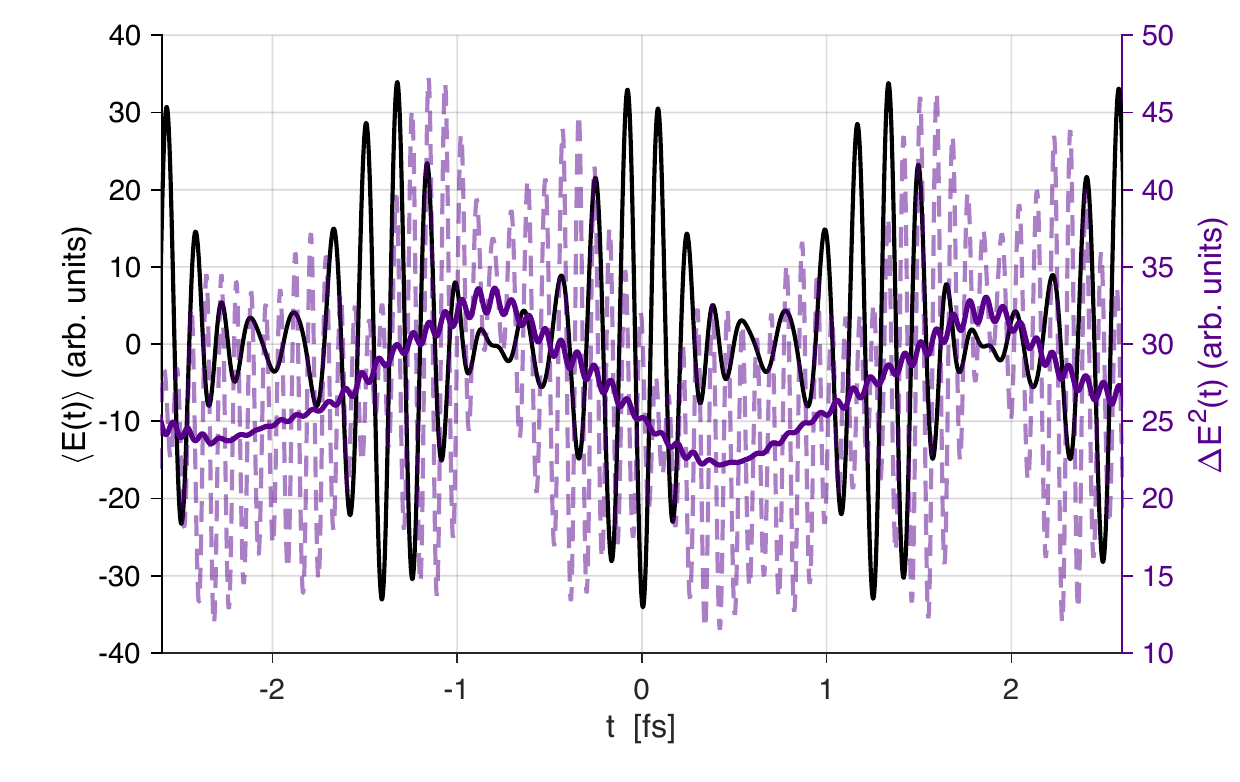}}
            \caption{\textbf{Time-domain reconstruction of attosecond emission.}  Reconstructed attosecond waveforms of the two-color HHG source showing the mean electric field $\langle E(t)\rangle$ (black), the instantaneous variance $\Delta E^2(t)$ (dashed purple), and its smoothed envelope (solid purple). The rapid sub-cycle oscillations of $\Delta E^2(t)$ reveal attosecond-scale noise bursts --direct evidence of quantum noise upconversion from the driving squeezed light into the XUV regime. The slower envelope follows the field-variance profile of the input bright squeezed vacuum, indicating that the underlying quantum noise structure is preserved through the high-harmonic generation process.}
			\label{fig:timeDomain}
		\end{center}
	\end{figure}

The coherent nature of the harmonics enables us to present a complementary time-domain perspective. At this stage, the attosecond pulse train emitted by the two-color HHG source is reconstructed on a shot-by-shot basis, unveiling the temporal evolution of its quantum noise. This approach allows us to preform a direct link between the spectral signatures of the harmonics and the time-resolved fluctuations of the emitted XUV field.
The reconstruction is performed by extracting the complex perturbation associated with each harmonic on a shot-by-shot basis via Eq. (1). The harmonics are then coherently linked by adding the unperturbed spectral phase, determined from the well-established atto-chirp extensively characterized over the past two decades \cite{paul2001observation,mairesse2005frequency}. This procedure yields an ensemble of attosecond waveforms, from which we extract the dynamical quantum fluctuations, quantified by the time-dependent variance $\Delta E^2(t)$. The resulting dynamics, together with the reconstructed mean value of the pulses, resolved on attosecond timescales, are shown in Figure ~\ref{fig:timeDomain}. 
The variance $\Delta E^2(t)$ (dashed purple line) exhibits rapid sub-cycle fluctuations that reveals the generation of attosecond-scale noise bursts in our experiment. These ultrafast field-fluctuations mark the direct time-domain manifestation of quantum noise upconversion, showing how the fluctuations of the driving squeezed light are transferred into the XUV regime. The quadrature asymmetry seen in Figure \ref{fig:Husimi} is manifested as oscillations of the variance at twice the frequency of the mean value oscillations. When averaged over these fine features, the smoothed variance (solid purple line) reveals a slowly varying envelope reminiscent of the input bright squeezed vacuum (BSV) field, indicating a signature of the driving quantum noise, which is preserved by the HHG mechanism. Scanning the two colors delay modulates the noise oscillations in phase and amplitude, yet maintains a slow temporal envelope mirroring the field variance of the input BSV.

\subsection*{Conclusion}
In this paper, we introduce the fundamental concepts of quantum light control into the attosecond XUV regime. Our study builds on three main milestones: controlling quantum fluctuations in XUV attosecond pulses, realizing optical tomography of quantum light in the XUV regime, and \emph{actively controlling the tunneling barrier} -- engineering and resolving its sub-cycle fluctuations. 
By combining infrared BSV with a strong coherent field, we imprint the squeezed fluctuations of the driving light onto both sub-cycle ionization dynamics and the emitted harmonics. Scanning the two-color delay provides precise control over the relative phase between photonic pathways, modulating both the mean intensity and the quantum state of the high harmonics. This interferometric scheme enables quantum state tomography -- long established in the visible and infrared -- for the first time in the XUV regime and on the attosecond timescale. We further report the first observation of optical tunneling driven by squeezed light, demonstrating that sub-cycle ionization fluctuations directly reflect the quantum correlations of the driving field. The tunneling statistics acquire a squeezed character, establishing a direct link between quantum optical noise and the tunneling mechanism.

Our results bridge quantum optics and attosecond science, opening new pathways for generating, characterizing, and controlling XUV pulses with non-classical properties. Non-classical attosecond physics holds great promise for advancing quantum-enhanced metrology, interferometry, and beyond, laying the foundation for attosecond-scale QED, where electron and photon quantum states can be manipulated with unprecedented temporal precision.

\clearpage 

%
\bibliography{science_template} 
\bibliographystyle{sciencemag}

%
%
%
%
%
%


\section*{Acknowledgments}
This project has received funding from the Israel Science Foundation (ISF) under grant agreements Nos. 1315/24 and 2626/23. M.E.T. gratefully acknowledges the support of the Council for Higher Education scholarship for excellence in quantum science and technology. N.D. is the incumbent of the Robin Chemers Neustein Professorial Chair. N.D. acknowledges the Minerva Foundation, the Israeli Science Foundation and the European Research Council for financial support. M.E.T., M.B., I.N., I.K., M.K., and  O.C. thank the Helen Diller Quantum Center for partial financial support. We thank Yossi Pilas for his technical support. 
\paragraph*{Author contributions:}
All authors contributed substantially to all aspects of the work.
\paragraph*{Competing interests:}
There are no competing interests to declare.
\paragraph*{Data and materials availability:}
The data supporting the findings of this study are available upon reasonable request.






\newpage


\renewcommand{\thefigure}{S\arabic{figure}}
\renewcommand{\thetable}{S\arabic{table}}
\renewcommand{\theequation}{S\arabic{equation}}
\renewcommand{\thepage}{S\arabic{page}}
\setcounter{figure}{0}
\setcounter{table}{0}
\setcounter{equation}{0}
\setcounter{page}{1} 


\begin{center}
\section*{Supplementary Materials for\\ \scititle}

\author{
	Matan~Even~Tzur$^{1*\dagger}$,
	Chen~Mor$^{2\dagger}$,
	Noa~Yaffe$^{2}$,
        Michael~Birk$^{1}$,
        Andrei~Rasputnyi$^{1,3,4}$,\and
        Omer~Kneller$^{2}$,
        Ido~Nisim$^{1}$, 
        Ido~Kaminer$^{1}$,
        Maria~Chekhova$^{1,3,4}$,
        Michael~Krüger$^{1}$,\and
        Misha Ivanov $^{1,5}$,
        Nirit~Dudovich$^{2}$,
        and Oren~Cohen$^{1}$
        \and
	\small$^{1}$Technion - Israel Institute of Technology, Haifa, Israel\and
	\small$^{2}$Department of Physics of Complex Systems, Weizmann Institute of Science, Rehovot, Israel\and
        \small$^{3}$Max Planck Institute for the Science of Light, Erlangen, Germany\and
                \small $^{4}$Friedrich-Alexander Universität Erlangen-Nürnberg
        \and 
        \small$^{5}$Max Born Institute, Berlin, Germany
        \and
	\small$^\ast$Corresponding author. Email: Matanev@campus.technion.ac.il\and
	\small$^\dagger$These authors contributed equally to this work.
}

\subsubsection*{This PDF file includes:}
Materials and Methods\\
Supplementary Text\\
Figures S1 to S5\\

\end{center}
\newpage


\subsection*{Materials and Methods}

\subsubsection*{Experimental setup}
\begin{figure}
    \centering
    \includegraphics[width=1\linewidth]{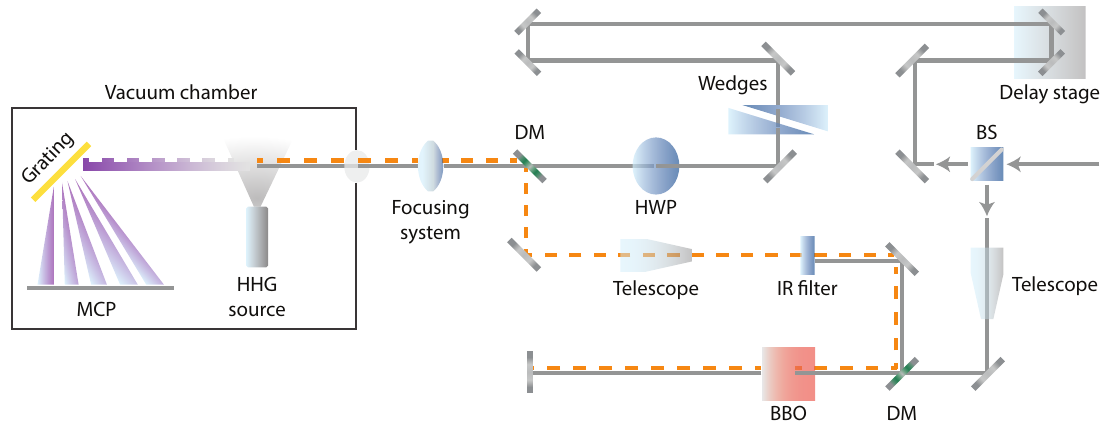}
    \caption{\textbf{Experimental setup.} 
    A strong IR pulse with 30fs duration and $800$nm central wavelength is split in a Mach-Zehnder interferometer into a BSV arm and a coherent arm.
    The BSV is generated by SPDC through a double pass in a BBO crystal with a central wavelength of $1600$nm.
    The relative delay between the IR and BSV is compensated using a delay stage and further controlled with sub-cycle precision using a pair of glass wedges located in the coherent arm. Polarization and focusing are separately manipulated to match the BSV and coherent modes.
    The two arms are combined using a dichroic mirror and focused into a continuous-flow glass nozzle filled with Kr to produce quantum perturbed HHG in the XUV regime.
}
    \label{fig:expSystem}
\end{figure}

A Ti:sapphire laser system operating at a 500Hz repetition rate delivers infrared (IR) pulses with $\sim 30$fs duration, at central wavelength of 800 nm and with an energy of $\sim 4.3$mJ per pulse.
The pulses are directed into a Mach–Zehnder interferometer, with a coherent state $800$nm arm and a bright-squeezed-vacuum (BSV) $1600$nm arm (see Figure \ref{fig:expSystem}).
The beam is split by a beamsplitter that allocates 90\% of the pulse energy to the delay-controlled coherent line and the remaining 10\% to the BSV line.
A phase-locked BSV with a central wavelength of 1600 nm is generated via spontaneous parametric down-conversion (SPDC) in a $3$mm, $\theta = 19.8^\circ$ type-I $\beta$-barium borate (BBO) crystal that can support pulses with a temporal duration of $15$fs.
The beam diameter is adjusted before the BBO using a beam-reducer telescope with a ratio of 1:3.
A $0^\circ$ mirror positioned $\sim 37$cm downstream from the BBO enhances BSV generation with a double pass through the crystal and filters out diverging BSV modes leading to the generation of BSV with an output energy of  $\sim 0.7\mu$J \cite{perez2014bright}. The properties of the generated BSV are described in detail in the next section.
A dichroic mirror is employed to reflect the BSV $1600$nm beam, while transmitting the coherent $800$nm beam.
An IR filter is used to further clear the BSV line from the fundamental coherent beam and the weak second harmonic generated in the BBO. Finally,  a beam-expander telescope enlarges the BSV beam diameter by a factor of 8 (to allow for tighter focusing of the BSV down the line).
In the coherent line, a delay line compensates for the relative optical paths to overlap the fundamental IR and the BSV field.
The sub-cycle delay between the IR and the BSV is controlled using a motorized pair of silica wedges, scanning the relative delay with a step size of $\sim200$as. 
A half-wave plate (HWP) rotates the IR field to align its polarization axis with the generated BSV.

The two lines are combined using a second dichroic mirror as the Mach-Zehnder beam combiner, and focused by a focusing system with an effective focal length of $\sim 1$m into a continuous-flow glass nozzle filled with noble gas atoms (Kr) to generate extreme-ultraviolet (XUV) attosecond pulses via high-harmonic generation (HHG). 

The XUV is spectrally resolved using a flat-field aberration-corrected concave grating, recorded using a micro-channel plate (MCP) detector and imaged by a charge-coupled device (CCD) camera.
The acquisition system is triggered by the laser to capture single-shot measurements with an integration time of $1$ms.

\subsubsection*{BSV generation and characterization}
We generate the BSV using two passes through the non-linear crystal, as described in the section above. This allows for sufficient amplification for the squeezed light to be bright, as well as for amplification of a single spatial and spectral dominant mode \cite{perez2014bright}.

\begin{figure}
		\centering
		\includegraphics[width=0.95\linewidth]{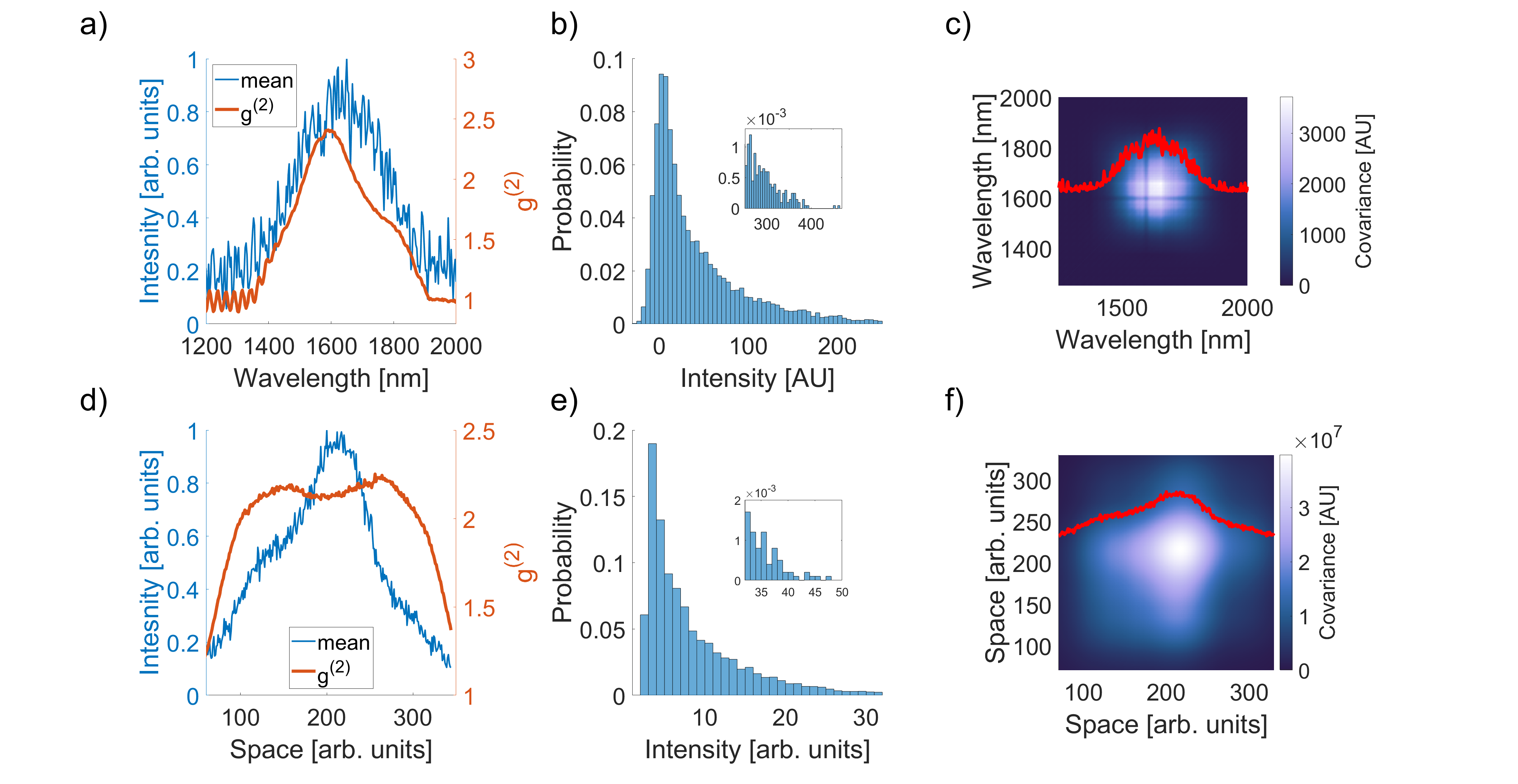    }
		\caption{ {\bf 1600 nm BSV source properties.} {\bf a),b),c)} Spectral properties. {\bf d),e),f)} Spatial properties.   {\bf a),d)} \gt\, and mean value on the spectrometer (camera).   {\bf b),e)} central wavelength (beam center) histogram. The inset continues the histogram on a different scale for higher photon numbers.	{\bf c),f)}	spectral (spatial) covariance. The red lines show the mean value measured in spectrum (space).}
			\label{fig:BSVsource}
	\end{figure}

To statistically characterize our BSV source we perform 10,001 single shot measurements on an InGaAs camera with a spectral range of $\sim 400 - 1650$nm and an InGaAs spectrometer with a spectral range of $\sim 890 - 2500 $nm. Figure \ref{fig:BSVsource} quantifies the statistical properties of the spatial and spectral modes of the BSV light. The mean value of the intensity  $I \propto \avg{\hat{N}}$ and second order correlation function$\gtm =(\frac{\avg{\hat{N^2}-\hat{N}}}{\avg{\hat{N}}^2})$ are presented in Fig \ref{fig:BSVsource}a,d. The data clearly demonstrates the BSV source exhibits superbunching (i.e., $\gtm>2$  \cite{mandel1995optical}). 
To extract \gt\, we used a softening parameter $\gtm = (\frac{\avg{\hat{N^2}}}{\avg{\hat{N}}^2+\epsilon_{soft}})$, to prevent \gt\, from diverging at the edges of the beam (due to background removal), while maintaining a change of $<1\%$ to \gt\, at the center of the beam. The deviation of \gt\, from the theoretical value of $\gtm = 3$ due to larger absorption of BSV photons by the SPDC crystal from brighter (higher photon number) realizations was already discussed and demonstrated elsewhere \cite{BSVreabsorption1976influence, BSVreabsorption1986photon}. 

Figure \ref{fig:BSVsource}b,e shows the raw histograms of the BSV intensity signal as measured by the spectrometer and the camera, in which the long tailed photon number distribution of the BSV (extending towards high photon numbers) is observed. The insets show the tails of the histograms at higher photon numbers, with axes scaled for visibility.  

Next, we consider the covariance of the BSV intensity signal, defined as the mutual variance of two observables (X,Y), $\Cov{X,Y} = \big\langle (X-\avg{X})(Y-\avg{Y}) \big\rangle$. 
Figure \ref{fig:BSVsource}c,f shows the covariance in space and spectrum. The covariance indicates that the BSV is nearly single mode in both space and spectrum as the entire beam is highly correlated. The effective number of modes was calculated by singular value decomposition \cite{law2000continuous} to be $\sim 1.2$.

\subsubsection*{XUV statistical data analysis}
We performed 10,001 single shot measurements for three different experimental configurations. The first includes the two-color experiment, performed with the strong coherent field and the BSV perturbation. The second is an experiment with a coherent drive only while the BSV is blocked. The third is a background measurement with both fields blocked. We then scanned the delay between the BSV and coherent fields in the overlap region, with 3,001 single shot measurements per delay. 

\subsubsection*{Single shot measurements}
To accurately obtain a full statistical analysis of light, we performed single shot measurements. We take the measurement with a micro-channel plate (MCP) detector with a phosphorous screen P43 which has a decay time of 1.5ms for 100\% to 10\% decay \cite{aase2011spatialMCP}. The laser shots are taken with a repetition rate of 500Hz and acquisition time of 1ms, triggering acquisition by the laser pulse. This setting resulted in effective $\sim 1.05$ pulses per measurement, due to leakage from pulses that arrived before the target pulse that did not fully decay yet. 

\subsubsection*{MCP response and \gt\, calibration}
\label{sec:g2Calib}
In this section we establish a calibration procedure to extract photon statistics using an MCP.
MCPs are commonly used in XUV spectrometers, as they have a high gain and have a relatively high efficiency in the XUV range. However, they introduce a statistical single photon response \cite{MCP_DISTRIBUTION_wiza1979microchannel,MCP_DISTRIBUTION_jones2019measuring,aase2011spatialMCP} even in the linear detection regime. We will now show the process that calibrates the detection to remove this statistical response of the MCP from the measurement, focusing on the mean value $\avg{N}$ and the normalized second moment \gt, but the process can be further generalized to higher moments.\\

\indent \textbf{Statistics of the light source}\\

The intensity of a quantum light source is proportional to the photon number which can be described by a stochastic variable $N$. 
By definition, \gt\, of the light source is:
\begin{equation}
    \gtm_{light} = \frac{\avg{{N}^2-{N}}}{\avg{{N}}^2} = 1+ \frac{\Var{{N}}}{\avg{{N}}^2}-\frac{1}{\avg{{N}}}
\end{equation}
In the high photon number limit:
\begin{equation}
    \gtm_{light} = 1+ \frac{\Var{{N}}}{\avg{{N}}^2}
\end{equation}
In the following we will consider this approximation for $\gtm_{light}$.\\

\indent \textbf{MCP single photon response}\\

For a single input photon, with no photon number uncertainty ($N=1$), the MCP statistical response introduces variance to the measurement. We denote the measured intensity by a random variable $X$. In this notation the dimensions of $X$ are $\frac{\text{measured intensity}}{\text{photon}}$, that is to statistically describe an intensity detector. Let us denote the mean of the MCP response as $\mu$ and the variance as $\sigma_{MCP}^2$. The second order normalized moment in this case is:
\begin{equation}
    \gtm_{1ph} \equiv 1+ \frac{ \sigma_{MCP}^2}{ {\mu^2}} 
\end{equation}

\indent \textbf{Full signal \gt}\\

The MCP measurements are given by a new random variable $I_m = \sum_{i=0}^N X_i$, with $X_i$ being the random variables representing the response of the MCP to the $i$th photon, signifying different random variables that share the same distribution. We note that for a detector with no uncertainty we have $I_m \propto N \Rightarrow \gtm_{detected} = \gtm_{light}$.\\

We proceed by examining the effect of the MCP on \gt.
The \gt \,measured by the MCP detector discussed above is:
\begin{equation}
    \gtm_{detected} = \frac{\avg{(\sum_{i=0}^N X_i)^2}}{\avg{\sum_{i=0}^N X_i}^2} 
    \label{eq:g2Det}
\end{equation}
In the linear detection regime the MCP responses to different photons are uncorrelated ($\Cov{X_i,X_j} = 0$ for $i\neq j$) and $X_i$ is independent of $N$. 
Since $X_i$ share the same distribution (the MCP single photon response) for all $i$, we can write the denominator of Eq. \ref{eq:g2Det} as:
\begin{equation}
    \avg{\sum_{i=0}^N X_i}^2 = \avg{\sum_{i=0}^N \avg{X_i}}_N^2 ={\avg{N X_i}^2}  = {\avg{N}^2 \avg{X_i}^2} = {\avg{N}^2 \mu^2}
\end{equation}

in which \(\mu=\langle X_i\rangle\) is independent of \(i\).
And the numerator as:
\begin{equation}
\avg{(\sum_{i=0}^N X_i)^2}= \avg{(N^2-N)X_i X_{j\neq i}+ N X_i^2} = \avg{N^2}\mu^2 +\avg{N} (\avg{X_i^2}-\mu^2) = \avg{N^2}\mu^2 + \avg{N} \sigma_{MCP}^2
\end{equation}
Therefore
\begin{equation}
    \gtm_{detected} = \frac{\avg{N^2}\mu^2 + \avg{N} \sigma_{MCP}^2}{ {\avg{N}^2 \mu^2}} 
\end{equation}
We define the quantity added to $\gtm_{detected}$ by the MCP for a single input photon ($N =1$) as  $G^{(2)}_{1ph} \equiv \frac{ \sigma_{MCP}^2}{ {\mu^2}} $ and finally we have:
\begin{equation}
    \gtm_{detected} =  \gtm_{light} + \frac{1}{\avg{N}} G^{(2)}_{1ph}
    \label{eq:MCPcalib}
\end{equation}

While $ G^{(2)}_{1ph}$ depends on the efficiency of the MCP and therefore can slightly change along the detector, $\gtm_{light}$ is fully independent of the detector or the photon number.\\

In our experiment we directly measure the intensity distribution $I_{detected}$. In the linear detection regime $\avg{I_{detected}}\propto \avg{I_{light}} \propto \avg{\hat{N}}$. From this distribution we can extract $\avg{I_{detected}}\propto \avg{\hat{N}}$ and $\gtm_{detected}$. We can then repeat the full statistical measurement, each time for a different value of $\avg{I_{detected}}$ and therefore of $\gtm_{detected}$.
We can fit these measurements  to Eq. \ref{eq:MCPcalib}, and obtain  $\gtm_{light}$.

\begin{figure}
		\centering
        \includegraphics[width=0.95\columnwidth]{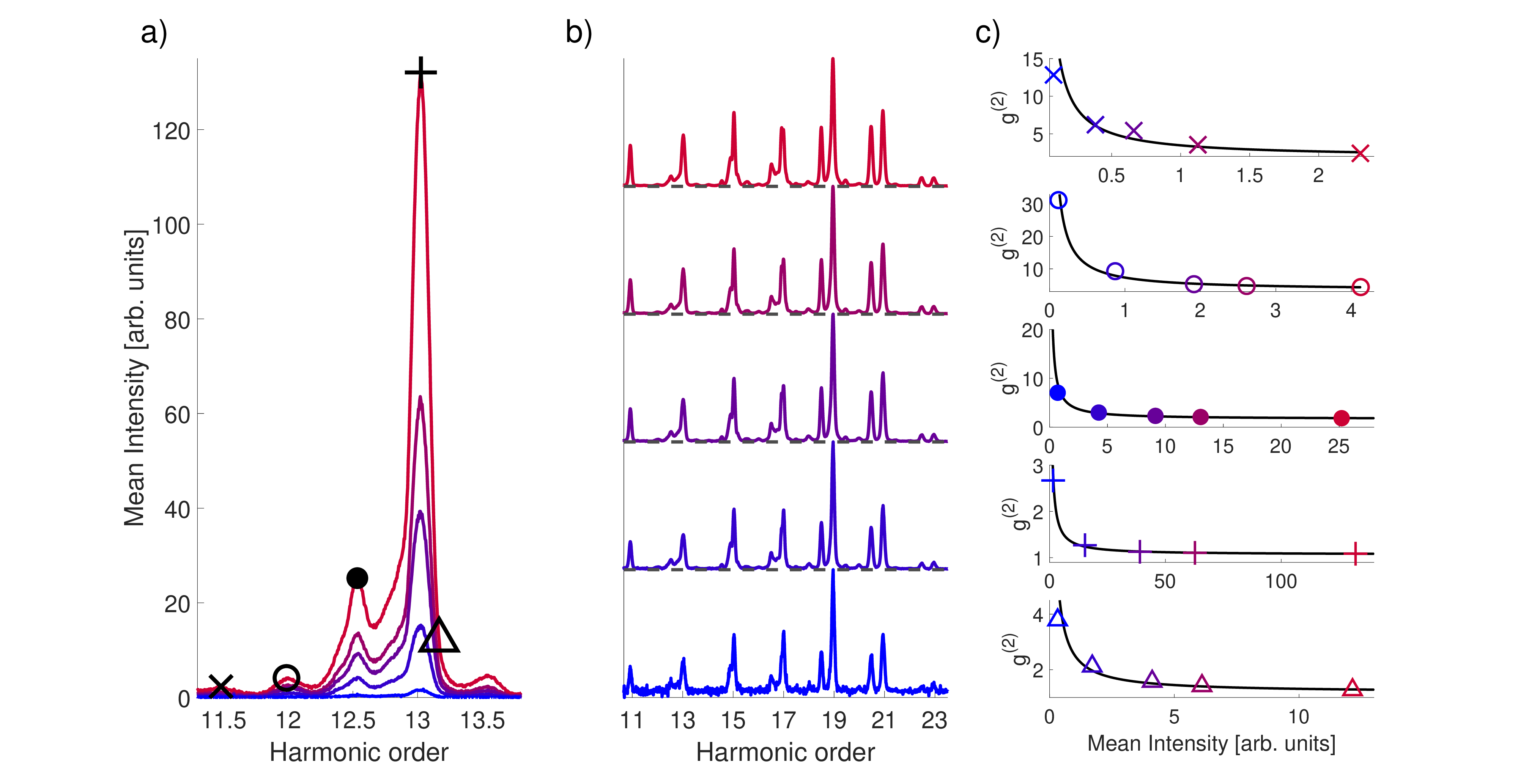    }
		\caption{ {\bf Verification of the MCP calibration procedure}    {\bf a)} Non-normalized spectra of different gas pressures, rising from blue to red, in a specific spectral region for clarity.  \bf{b)} Normalized mean spectrum for the same gas pressures with the same color code. It is apparent that the normalized spectra are identical up to SNR, indicating that scanning the gas pressure modifies the mean photon number only. {\bf c)} Fits of \gt as a function of intensity to Eq. \ref{eq:MCPcalib} at different spectral regions, seen as corresponding markers in { a)}. }
			\label{fig:MCPpressureCalib}
	\end{figure}

We verify this calibration by scanning the gas pressure, and therefore the number of emitters. In a dilute gas the average photon number $\avg{\hat{N}}$ scales linearly with the number of emitters, while $\gtm_{light}$, remains unchanged. Figure \ref{fig:MCPpressureCalib} describes this calibration, demonstrating its validity. Figure \ref{fig:MCPpressureCalib}a describes the measured spectra at different pressure values. Figure \ref{fig:MCPpressureCalib}b shows the spectra of different pressures (normalized for visibility), from the lowest pressure (blue) to the highest pressure (red). Their similarity substantiates that changing the pressure only scales the photon number,while keeping the normalized moments of the light itself such as $\gtm_{light}$ unchanged. Finally, Figure \ref{fig:MCPpressureCalib}c describes the measured  \gt at different points in the spectra, shown as markers on \ref{fig:MCPpressureCalib})a, together with the fit to Eq. \ref{eq:MCPcalib}. It is clear that a good agreement between the measured values and the theoretical fit is achieved even in very low flux regions of the spectrum. 

In our experiment we perform a similar calibration directly by comparing different pixels on the detector along a spectral line, representing different number of emitters.

\subsubsection*{Background subtraction}
The statistical properties of the measurement are dictated by the statistics of the signal as well as the statistics of any background:
\begin{equation}
    M = S + B
\end{equation}
Where $M$ is the measurement, $S$ is the signal (including the MCP noise discussed above), $B$ is the background noise (present without input XUV light), and all three parameters represent random variables. 

Next, we expand the moments of $S$ in terms of the moments of $B$ and $M$. The mean value trivially decomposes to \(\langle M \rangle=\langle S \rangle+\langle B \rangle\) while the variance is given by:

\begin{equation}
    \Var{M}  = \Var{S+B} = \avg{S^2} +\avg{B^2} +  2 \avg{S} \avg{B} - \avg{S}^2 - \avg{B}^2  - 2 \avg{S} \avg{B} =  \Var{S} + \Var{B}
\end{equation}
Where we have used the lack of correlations between the background noise and the signal.

Finally, we have the variance of the signal:
\begin{equation}
    \Var{S}  =  \Var{M} - \Var{B}
\end{equation}

Therefore, to obtain the mean value and \gt\, of the signal, the mean and variance of the background signal are subtracted from the measurement.

In our experiment we had two types of background mechanisms. The first, originating from the detection system, is measured by removing the gas from the beam path, fully eliminating the HHG radiation. The second originates from leakage between neighboring regions on the MCP resulting in signal from the bright odd harmonics affecting the weak half integer harmonics and the even harmonics. Therefore, the background associated with the odd harmonics can be removed by recording the signal in the absence of XUV radiation. The background associated with the half integer and even harmonics is removed by measuring the HHG signal while blocking the BSV source. Equation \ref{eq:MCPBG} describes \gt\, in the presence of background with a mean value $\avg{N_{BG}}$, and second order normalized coherence $\gtm_{BG}$:

\begin{equation}
    \gtm_{detected} = \frac{\avg{(N+N_{BG})^2}+ \avg{N+N_{BG}} \sigma^2/\mu^2 }{{(\avg{N}+\avg{N_{BG}})^2 }} =  \frac{\gtm_{light}\cdot\avg{N}^2+\gtm_{BG}\cdot\avg{N_{BG}}^2+ \avg{N+N_{BG}} \sigma^2/\mu^2 }{{(\avg{N}+\avg{N_{BG}})^2 }}
    \label{eq:MCPBG}
\end{equation}

 Clearly, when $\avg{N}$ grows with respect to $\avg{N_{BG}}$, \gt converges to $\gtm_{light}$.
After removing the full background signal we obtain the \gt\, calibration curve from Eq. \ref{eq:MCPcalib}.


\subsection*{Derivation of Eq. (1) in the main text}

In this section, we derive Eq. 1 in the main text, describing harmonic intensities in our experiment, as measured on a shot-by-shot basis. We employ the semi-classical strong-field approximation theory\cite{lewenstein1994theory}. For an atom with an ionization potential $I_p$, and an electron with a mass $m$, that interacts with a linearly polarized light, the semi-classical action is given by:  

\begin{equation}
S = \int_{t'}^{t} dt'' \left( \frac{\left(p - eA(t'')\right)^2}{2m} + I_p \right)
\end{equation}
\begin{equation}
A (t) = A_0 sin(\omega t)+A_p sin(\omega t/2 +\phi_{TC}) 
\end{equation}

in which $\omega$ corresponds to the angular frequency of the bright 800 nm coherent beam, and $\phi_{\text{TC}}$ is the two-color phase. $A_0$ and $A_p$ are the vector potential amplitudes of the 800 nm pump and 1600 nm perturbation, respectively. Assuming $A_p \ll A_0$, this can be written as:  

\begin{equation}
S \approx S_0 + \sigma
\end{equation}

In which  

\begin{equation}
S_0 = \int_{t'}^{t} dt'' \left( \frac{\left(p - eA_0 \sin(\omega t'')\right)^2}{2m} + I_p \right)
\end{equation}

\begin{equation}
\sigma = -eA_p \int_{t'}^{t} dt'' \left( \frac{\left(p - eA_0 \sin(\omega t'')\right) \sin\left(\frac{\omega t''}{2} + \phi_{\text{TC}}\right)}{m} \right)
\end{equation}

The $\Omega$ frequency component of the electronic dipole moment is given by a sum of four contributions from four consecutive half cycles, in a close analogy to previous two color experiments \cite{dudovich2006measuring,dahlstrom2011quantum}:

\begin{align}
X_\Omega = \sum_n x_\Omega^{(n)} \bigg\{ 
& e^{i\sigma (t', t, p)} 
- e^{i\sigma (t' + \frac{T}{2}, t + \frac{T}{2}, -p) - i\Omega \frac{T }{2}} \notag + e^{i\sigma (t' + T, t + T, p) - i\Omega T} \notag \\
& - e^{i\sigma (t' + \frac{3T}{2}, t + \frac{3T}{2}, -p) - i\Omega \frac{3T}{2}} 
\bigg\}
\end{align}

in which $T = 2\pi / \omega$, the summation over $n = 1, 2$ corresponds to summation over short ($n = 1$) and long ($n = 2$) trajectories, and $x_\Omega^{(n)}$ is the unperturbed ($A_p = 0$) dipole moment from a single-half cycle at frequency $\Omega$ . 
The complex phase perturbation to the electron's action satisfies the following relations:
\begin{equation}
\begin{gathered}
\sigma_1(n,\Omega,\phi_{\text{TC}}) = \sigma(t',t,p) =\\
 -eA_p \int_{t'^{(n,\Omega)}}^{t^{(n,\Omega)}} dt'' \left[ p^{(n,\Omega)} - eA_0 \sin(\omega t'') \right] 
\quad \times \frac{\sin\left( \frac{\omega t''}{2} + \phi_{\text{TC}} \right)}{m} 
= \frac{d\sigma_2}{d\phi_{\text{TC}}}
\end{gathered}
\label{eq:sigma1}
\end{equation}

\begin{equation}
\begin{gathered}
\sigma_2(n,\Omega,\phi_{\text{TC}}) = \sigma_1\left(t' + \frac{T}{2}, t + \frac{T}{2}, -p, \phi_{\text{TC}}\right) = \\
 eA_p \int_{t'^{(n,\Omega)}}^{t^{(n,\Omega)}} dt'' \left[ p^{(n,\Omega)} - eA_0 \sin(\omega t'') \right] \times \frac{\cos\left( \frac{\omega t''}{2} + \phi_{\text{TC}} \right)}{m} 
= -\frac{d\sigma_1}{d\phi_{\text{TC}}}
\end{gathered}
\label{eq:sigma2}
\end{equation}

\noindent We note that 
\[
\sigma\left(t' + T, t + T, -p\right) = -\sigma_1(n, \Omega, \phi_{\text{TC}})
\]
and
\[
\sigma\left(t' + \frac{3T}{2}, t + \frac{3T}{2}, -p\right) = -\sigma_2(n, \Omega, \phi_{\text{TC}}).
\]
Therefore, we may reorganize the expression \( X_\Omega \) depending on the value of \( \Omega \) :

\begin{equation}
X_\Omega = 2x_\Omega \times 
\begin{cases} 
\cos(\sigma_{1 }) + \cos(\sigma_{2 }) & \text{if } \Omega = (2m -1)\omega, \\ 
\cos(\sigma_{1 }) - \cos(\sigma_{2 }) & \text{if } \Omega = 2m\omega, \\ 
i\sin(\sigma_{1 }) + \sin(\sigma_{2 }) & \text{if } \Omega = \left(2m - \frac{3}{2}\right)\omega, \\ 
i\sin(\sigma_{1 }) - \sin(\sigma_{2 }) & \text{if } \Omega = \left(2m - \frac{1}{2}\right)\omega.
\end{cases}
\label{eq:sfa}
\end{equation}
in which \(m\) is an integer. This constitutes Eq. 1 in the main text. We note that the perturbations to the electron's action, $\sigma_{1,2}$, are complex valued. Their real and imaginary parts are denoted by $\sigma_{1,2}\equiv\alpha_{1,2}+i\beta_{1,2}$, where $\alpha_j$ and $\beta_j$ correspond to the phase and amplitude of the emission from half cycle $j$, respectively. 

\subsection*{Reconstruction of $(\sigma_1,\sigma_2)$ from measured harmonic intensities}
We extract the perturbation parameters \(\sigma_{1,2}=\alpha_{1,2}+i\beta_{1,2}\), by fitting a quartet of measured harmonic-intensity families to Eq.(1) in the main text. Based on Eq. 1, we can express the ratio between the BSV perturbed harmonics and the unperturbed harmonics, in terms of \(\alpha_i,\beta_i\): 

\begin{equation}
\begin{gathered}
A \equiv \frac{|X_{(2m-1)\omega}|^2}{|x_{(2m-1)\omega}|^2} = \left(\cos(\alpha_1)\cosh(\beta_1) + \cos(\alpha_2)\cosh(\beta_2)\right)^2 \\
+ \left(\sin(\alpha_1)\sinh(\beta_1) + \sin(\alpha_2)\sinh(\beta_2)\right)^2
\end{gathered}
\end{equation}
\begin{equation}
\begin{gathered}
B \equiv \frac{|X_{2m\omega}|^2}{|x_{(2m-1)\omega}|^2} = \left(\cos(\alpha_1)\cosh(\beta_1) - \cos(\alpha_2)\cosh(\beta_2)\right)^2 \\
+ \left(\sin(\alpha_1)\sinh(\beta_1) - \sin(\alpha_2)\sinh(\beta_2)\right)^2
\end{gathered}
\end{equation}
\begin{equation}
\begin{gathered}
C \equiv \frac{|X_{(2m-1/2)\omega}|^2}{|x_{(2m-1)\omega}|^2} = \left(-\cos(\alpha_1)\sinh(\beta_1) + \sin(\alpha_2)\cosh(\beta_2)\right)^2 \\
+ \left(\sin(\alpha_1)\cosh(\beta_1) + \cos(\alpha_2)\sinh(\beta_2)\right)^2
\end{gathered}
\end{equation}
\begin{equation}
\begin{gathered}
D \equiv \frac{|X_{(2m-3/2)\omega}|^2}{|x_{(2m-1)\omega}|^2} = \left(-\cos(\alpha_1)\sinh(\beta_1) - \sin(\alpha_2)\cosh(\beta_2)\right)^2 \\
+ \left(\sin(\alpha_1)\cosh(\beta_1) - \cos(\alpha_2)\sinh(\beta_2)\right)^2
\end{gathered}
\end{equation}
In which \(|X_{q\omega}|^2\) is the intensity of the \(q\omega\) harmonic in the presence of the BSV, and \(4|x_{(2m-1)\omega}|^2\) is the intensity of the BSV-unperturbed emission of the adjacent odd harmonic (\(x_{(2m-1)\omega}\) is proportional to the unperturbed dipole moment at a frequency \((2m-1)\omega\) generated by a single half-cycle of the driving field; see Equation \eqref{eq:sfa}).  

Extracting the perturbation parameters \(\alpha_i,\beta_i\) requires a careful analysis on a pixel-by-pixel level. To recover these parameters we employ the following datasets: (i) coherent-only driving field (``Coh'', odd only harmonics), (ii) background measurements, ``BG'', no XUV radiation, and (iii) coherent driving field with bright squeezed vacuum perturbation (``BSV'', full emission spectrum perturbed by BSV). 

For each two-color delay \(\tau\) and each harmonic order, we extract the following signals:
\[
S^{\rm (A)}=S_{BSV}-\overline{S}_{\rm BG},\qquad
S^{\rm (B)}=S_{BSV}-\overline{S}_{\rm Coh}.
\]
in which \(S\) is a single intensity measurement of the harmonic (perturbed by the BSV), \(\overline{S}_{\rm BG}\) is the mean intensity measured on the same detector pixels when both the coherent and BSV beams are off, and  \(\overline{S}_{\rm Coh}\) is the mean intensity measured when the BSV beam is off, while the coherent beam is on. 

For each \emph{odd} harmonic order $N$ (e.g.\ $\{13,15,17,19,21\}$) and time delay \(\tau\), we assemble a quartet of harmonic intensities:
\begin{align}
\text{A\((N,\tau)\) (odd)}: &\quad q=N \ \text{from}\ S^{\rm (A)} \ ,\\
\text{B\((N,\tau)\) (even)}: &\quad q=N+1 \ \text{from}\ S^{\rm (B)} \ ,\\
\text{C\((N,\tau)\) (odd+}\tfrac{1}{2}\text{)}: &\quad q=N-\tfrac{1}{2} \ \text{from}\ S^{\rm (B)} \  ,\\
\text{D\((N,\tau)\) (odd+}\tfrac{3}{2}\text{)}: &\quad q=N+\tfrac{1}{2} \ \text{from}\ S^{\rm (B)} \ .
\end{align}
Here \(S^{(A)}\) denotes the  BSV-perturbed odd-harmonic intensity measured in the after background subtraction, whereas \(S^{(B)}\) denotes the BSV-perturbed intensity measured after coherent-baseline subtraction. Using the coherent-subtracted data \(S^{(B)}\) for even and half-integer harmonics removes residual spectral leakage from bright odd harmonics into neighboring bins and ensures that the extracted perturbation parameters \(\sigma_{1,2}\) vanish when the BSV is blocked. 

After assembling the quartet of background and coherent subtracted reduced intensities, each single-shot intensity quartet \(A,B,C,D(N,\tau)\) is normalized with a common normalization factor \(Norm(N,\tau)\): 
\begin{equation}
A(N,\tau)\rightarrow\frac{\,4A(N,\tau)}{\mathrm{Norm}(N,\tau)},\quad
B(N,\tau)\rightarrow\frac{\,4B(N{+}1,\tau)}{\mathrm{Norm}(N,\tau)},\quad
C(N,\tau)\rightarrow\frac{\,4C(N{-}\tfrac{1}{2},\tau)}{\mathrm{Norm}(N,\tau)},\quad
D(N,\tau)\rightarrow\frac{\,4D(N{+}\tfrac{1}{2},\tau)}{\mathrm{Norm}(N,\tau)}.
    \label{eq:norm-intensities}
\end{equation}
The normalization coefficient is chosen as the shot-average of \(A(N,\tau)\).  The normalized values of \(A,B,C,D(N,\tau)\) are the values used for the reconstruction of the perturbation parameters for the harmonic quartet adjecent to the odd harmonic \(N\) at a time delay \(\tau\), i.e., from them we extract the parameters \(\sigma_{1,2}(N,\tau)\) on a shot by shot basis. The factor of 4 in the normalization is chosen such that, when the intensities of the even and half-integer harmonics vanish, the normalized odd-harmonic intensity equals 4. This convention ensures consistency with the normalization used in Eq.~\eqref{eq:sfa}. 

\subsubsection*{Parametric model and unknowns}
In the following stage we reconstruct the complex parameters by fitting the model predictions from Equation~\eqref{eq:sfa} (corresponding to Equation~1 in the main text), to the measured (normalized) harmonics intensities, extracted from Equation~\eqref{eq:norm-intensities}. The model describes the harmonic intensities as functions of the complex perturbation parameters $\sigma_{1,2}$ and the BSV-unperturbed odd-harmonic emission amplitude, given by $4|x_{(2m-1)\omega}|^2$. 
Before applying this analysis we note that irrespective of the presence of the BSV perturbation, the unperturbed odd-harmonic intensity $4|x_{(2m-1)\omega}|^2$ itself fluctuates from shot to shot (e.g., due to fluctuations of the coherent \(800\) nm beam). If these fluctuations are ignored, they would be falsely attributed to variations in the perturbation parameters $\sigma_{1,2}$. To avoid this cross-contamination, we introduce a per-shot normalized scaling factor $X_2$ proportional to the unperturbed emission amplitude 
\begin{equation}
X_2= |x_{(2m-1)\omega}|^2 / \mathrm{Norm}(N,\tau)
    \label{eq:X2}
\end{equation}
This term accounts for intrinsic fluctuations of the coherent HHG emission and is fitted together with $\sigma_{1,2}$ for each shot.

To reconstruct the parameters \(\sigma_{1,2}\) and the scaling factor \(X_2\), we compare the measured normalized harmonic intensities \(A_N^{\mathrm{(meas)}},\dots,D_N^{\mathrm{(meas)}}\) with their model counterparts derived from Eq.~\eqref{eq:sfa}. For compactness, we write
\[
\begin{aligned}
A_N^{\mathrm{(model)}} &= X_2\,|\cos\sigma_1+\cos\sigma_2|^2,\\
B_N^{\mathrm{(model)}} &= X_2\,|\cos\sigma_1-\cos\sigma_2|^2,\\
C_N^{\mathrm{(model)}} &= X_2\,|\,i\sin\sigma_1+\sin\sigma_2|^2,\\
D_N^{\mathrm{(model)}} &= X_2\,|\,i\sin\sigma_1-\sin\sigma_2|^2.
\end{aligned}
\]
The five fitting parameters are \(\alpha_{1,2},\beta_{1,2}\) (with \(\sigma_i=\alpha_i+i\beta_i\)) and \(X_2>0\), while the measured values remain fixed. The residual vector between measurement and model is
\[
\mathbf{r}=
\begin{bmatrix}
A_N^{\mathrm{(meas)}}-A_N^{\mathrm{(model)}}\\
B_N^{\mathrm{(meas)}}-B_N^{\mathrm{(model)}}\\
C_N^{\mathrm{(meas)}}-C_N^{\mathrm{(model)}}\\
D_N^{\mathrm{(meas)}}-D_N^{\mathrm{(model)}}
\end{bmatrix},
\qquad
\mathbf{F}=
\begin{bmatrix}
r_1/A_N^{\mathrm{(meas)}},\,
r_2/B_N^{\mathrm{(meas)}},\,
r_3/C_N^{\mathrm{(meas)}},\,
r_4/D_N^{\mathrm{(meas)}}
\end{bmatrix}^{\!\!T}.
\]
Normalizing the residuals in this way is essential: without it, the harmonic with the largest intensity would dominate the fit, causing the optimization to minimize its residual at the expense of the others. A nonlinear least-squares solver (\texttt{lsqnonlin}, MATLAB) iteratively updates \(\alpha_{1,2},\beta_{1,2},X_2\) to minimize the normalized residual norm \(\|\mathbf{F}\|^2\), subject to
\[
\alpha_{1,2}\in[-\pi,\pi],\quad
\beta_{1,2}\in[-\pi,\pi],\quad
X_2>0.
\]

\subsubsection*{Two-stage alternating solver}

Because of the intrinsic symmetries of Eq.~(1) in the main text, reconstructing \(\sigma_{1,2}\) from the measured harmonic intensities is not unique. In particular, the intensities are invariant under the transformations \(\sigma_{1,2}\rightarrow-\sigma_{1,2}\) and \((\sigma_1,\sigma_2)\leftrightarrow(\sigma_2,-\sigma_2)\). As a result, several mathematically valid solutions correspond to the same physical data. Although each shot must physically correspond to a single physical pair \((\sigma_1,\sigma_2)\), numerical solvers may converge to any of these symmetric branches. 
To remove this ambiguity, we adopt a two-stage alternating strategy. In the first stage, we intentionally break the symmetry to guide the solver toward a single branch; in the second stage, we reintroduce the original equations and refine the result to the physical solution.

\paragraph{Stage 1: solve modified (symmetry-broken) equations to obtain an initial guess.}
Here we temporarily alter the trigonometric combinations in Equation \eqref{eq:sfa} by replacing
\[
\cos \sigma_1 \pm \cos \sigma_2 \;\longrightarrow\; \cos \sigma_1 \pm 2\cos \sigma_2,
\qquad
i\sin \sigma_1 \pm \sin \sigma_2 \;\longrightarrow\; i\sin \sigma_1 \pm 2\sin \sigma_2,
\]
which breaks the degeneracy between equivalent solutions. The system is then solved iteratively, starting from a random initial guess for \((X_2,\alpha_1,\alpha_2,\beta_1,\beta_2)\), and iteratively:
\begin{enumerate}
\item fixing \(X_2\) and solving for \((\alpha_1,\alpha_2,\beta_1,\beta_2)\), and
\item fixing \((\alpha_1,\alpha_2,\beta_1,\beta_2)\) and solving for \(X_2\).
\end{enumerate}
These two steps are repeated until convergence. The resulting solution provides a consistent, symmetry-broken initialization for Stage~2.

\paragraph{Stage 2: solve the physical equations using the Stage~1 result as an initial condition.}
With the initial guess supplied by Stage~1, we repeat the same alternating minimization using the unmodified, physical equations (Equation \eqref{eq:sfa}). Because the initialization already selects one symmetry branch, the solver now converges reliably to the unique physical solution rather than to an arbitrary symmetric equivalent.

\subsection*{Experimental Husimi reconstruction from $(\alpha_{1,2},\beta_{1,2},X_2)$}

We reconstruct experimental Husimi (Q) distributions for selected harmonic families using the fitted parameters
\[
\alpha_1,\ \alpha_2,\ \beta_1,\ \beta_2,X_2.
\]
This reconstruction maps the extracted perturbation parameters onto a phase-space representation of the emitted field. We define quantity \(x_\Omega\equiv\sqrt{X_2}\) which represents the magnitude of the unperturbed harmonic field amplitude, since \(X_2\) is proportional to its intensity (see Eqs.~\eqref{eq:X2},~\eqref{eq:sfa}). When the BSV perturbation is blocked, \(x_\Omega\) is proportional to the per-shot coherent XUV amplitude.

Because the optical phase of the unperturbed emission cannot be measured on a shot-by-shot basis, we treat it as a random variable with uniform distribution \(\phi\sim\mathrm{Unif}[0,2\pi)\). Each shot is therefore assigned a complex unperturbed amplitude
\[
E_0 \equiv (x_\Omega-\langle X_\Omega\rangle)\,e^{i\phi} + \langle X_\Omega\rangle,
\]
where \(\langle X_\Omega\rangle\) is the ensemble mean over shots at the same two-color delay and harmonic order. This assumption corresponds to a coherent XUV state with amplitude noise distributed equally between its two quadratures \cite{Gariepy2014,Fleischer2014,huang2018polarization}.

\paragraph{Normalization.}
To express all quadratures in units of the vacuum noise level of the unperturbed coherent state, we rescale the reconstructed amplitudes such that \(\langle \Delta X^2 \rangle = \langle \Delta P^2 \rangle = 1/2\). The quadrature variances of the unperturbed amplitudes \(E_0\) are
\[
\sigma_X^2=\mathrm{Var}\big(\Re E_0\big),\qquad
\sigma_P^2=\mathrm{Var}\big(\Im E_0\big),
\]
and the corresponding normalization factor is
\[
\sigma_{qp}=\sqrt{\tfrac{1}{2}(\sigma_X^2+\sigma_P^2)},\qquad
\mathcal{N}=\sqrt{2}\,\sigma_{qp}.
\]
Dividing all reconstructed quadrature amplitudes by \(\mathcal{N}\) ensures that the Husimi distributions are expressed in canonical quantum-optical units, with quadrature variance \(1/2\) when \(\sigma_{1,2}=0\) (i.e., in the BSV-unperturbed case). 

\paragraph{Husimi reconstruction.}
For each shot, the harmonic quadrature amplitudes are then obtained by combining the unperturbed amplitude \(E_0\) with the complex perturbations \(\sigma_{1,2}\) according to the four harmonic families:
\begin{align}
\text{Row 1 (odd-like)}&:\quad Q = \frac{E_0\,[\cos \sigma_1 + \cos \sigma_2]}{\mathcal{N}},\\
\text{Row 2 (even-like)}&:\quad Q = \frac{E_0\,[\cos \sigma_1 - \cos \sigma_2]}{\mathcal{N}},\\
\text{Row 3 (odd+}\tfrac{1}{2}\text{)}&:\quad Q = \frac{E_0\,[\,\sin \sigma_2 + i\,\sin \sigma_1\,]}{\mathcal{N}},\\
\text{Row 4 (odd+}\tfrac{3}{2}\text{)}&:\quad Q = \frac{E_0\,[\,-\sin \sigma_2 + i\,\sin \sigma_1\,]}{\mathcal{N}}.
\end{align}
The corresponding normalized quadratures are \(X=\Re Q\) and \(P=\Im Q\). From the ensemble of \(N\) laser shots, the set of quadratures \(\{(X_j,P_j)\}_{j=1}^N\) yields the experimental Husimi distribution \(Q(X,P)\), which represents the probability density to measure a coherent state \(\alpha=X+iP\)  with quadrature amplitudes \((X,P)\).

\subsection*{Wigner reconstruction of the single-photon channel of half-integer harmonics}

The generation of half-integer harmonics can be described perturbatively with respect to the BSV field. This field can be viewed as a superposition of two channels associated with first- and third-order perturbations, corresponding to the absorption or emission of one or three BSV photons during the strong-field interaction. In the following, we isolate and reconstruct the Wigner function of the single-photon channel, which corresponds to the absorption or emission of a single BSV photon during the strong-field process. 

Starting from Eq.~\eqref{eq:sfa}, the complex field of the $(2m-\tfrac{3}{2})\omega$ harmonic is
\[
X_{(2m-\tfrac{3}{2})\omega} \ \propto\ i\sin\sigma_1 + \sin\sigma_2,
\]
with $\sigma_j=\alpha_j+i\beta_j$. Expanding this expression in powers of $\sigma_{1,2}$ decomposes the response into distinct photonic pathways:
\[
X_{(2m-\tfrac{3}{2})\omega}(\phi_{\mathrm{TC}})
\ \propto\ 
\underbrace{\big(\sigma_2+i\sigma_1\big)}_{\text{single BSV photon}}
\ -\ \tfrac{1}{6}\,\underbrace{\big(\sigma_2^3+i\sigma_1^3\big)}_{\text{three BSV photons}}
\ +\ \ldots
\]
The first term corresponds to the single-photon channel, for which we define the complex field amplitude
\[
\epsilon \equiv \mathcal{X}+i\mathcal{P} = \sigma_2(0) + i\,\sigma_1(0).
\]
When the two-color delay \(\phi_{\mathrm{TC}}\) is scanned, Eqs.~\eqref{eq:sigma1}–\eqref{eq:sigma2} describe a rotation of $(\sigma_1,\sigma_2)$ in the complex plane, and thus of \(E\):
\[
\epsilon(\phi_{\mathrm{TC}})=\epsilon\,e^{-i\phi_{\mathrm{TC}}}, 
\qquad 
X_\phi=\Re\!\big\{\epsilon\,e^{-i\phi_{\mathrm{TC}}}\big\}.
\]
In this picture, $\phi_{\mathrm{TC}}$ plays the role of the local-oscillator phase in homodyne detection: varying $\phi_{\mathrm{TC}}$ samples different quadrature projections \(X_\phi\), yielding the set of marginals \(P(X_\phi)\) associated with the single-photon mode \(\epsilon\). The Wigner function \(W(X,P)\) of this mode is then obtained—up to an overall scale—by performing an inverse Radon transform following standard homodyne-tomography protocols \cite{Smithey1993,scully1997quantum}.

Figure~\ref{fig:wigner_single_photon} shows the reconstructed Wigner function for \(q=14.5\). The measured quadrature \(X_\phi(\phi_{\mathrm{TC}})\) exhibits negligible mean displacement and a phase-dependent variance, with \(\min_\phi \Delta X_\phi^2 / \max_\phi \Delta X_\phi^2 \simeq 0.55\). The corresponding \(W(X,P)\) is centered near the origin and elongated along one quadrature axis, consistent with a squeezed single-photon channel (Fig.~\ref{fig:wigner_single_photon}a–b).

\subsection*{TDSE Simulations and Husimi Reconstruction}

In this section, we outline the numerical methodology used to calculate the quantum states of harmonics presented in Figure 4 of the main text. The procedure consists of two steps. First, we solve the semi-classical time-dependent Schrödinger equation (TDSE) for an atom driven by a classical, bi-chromatic, \(\omega-\omega/2\) laser field, sampling a broad range of the \(\omega/2\)-field amplitudes and phases. 
Second, we evaluate the quantum state of the harmonics (represented by the Husimi distribution) using the joint probability density function of their amplitudes and phases.

\subsubsection*{1. Semi-classical Time-Dependent Schrödinger Equation (TDSE) simulations}  

We solve the one-dimensional time-dependent Schrödinger equation (TDSE)  

\[
i \hbar \,\frac{\partial}{\partial t}\Psi(x,t) = \hat{H}(t)\Psi(x,t),
\] 
with Hamiltonian  

\[
\hat{H}(t) = -\frac{\hbar^2}{2m_e}\frac{\partial^2}{\partial x^2} + V_{\mathrm{atom}}(x) + V_{\mathrm{abs}}(x) + xE(t).
\]  We use atomic units throughout. 

\paragraph{Atomic potential (Krypton).}  
The krypton atom is modeled by a short-range Pöschl–Teller potential  

\[
V_{\mathrm{atom}}(x) = -\frac{V_0}{\cosh^2(\alpha x)},
\]

with parameters  

\[
V_0 = 0.855, \qquad \alpha = 0.68,
\]

chosen to exhibit the ionization potential of Kr.   

\paragraph{Numerical box and absorber.}  
The TDSE is solved over a discrete spatial grid in one dimension, centered at \(x=0\), with a uniform grid spacing \(dx\) and grid size \(L_x\): 
\[
L_x = 101 \ \text{a.u.}, \qquad dx = 0.12 \ \text{a.u.},
\] To suppress reflections from the grid edge, a complex absorbing potential is applied for \(|x| > R_{\mathrm{abs}}\):  

\[
V_{\mathrm{abs}}(x) = -i \, \mathrm{AbsStrength}\, \big(|x|-R_{\mathrm{abs}}\big)^3 \, \Theta\!\big(|x|-R_{\mathrm{abs}}\big),
\]  

where  

\[
R_{\mathrm{abs}} = 0.7\frac{E_0}{\omega^2}, 
\qquad \mathrm{AbsStrength} = 5\times 10^{-2}.
\] in which \(E_0\) and \(\omega\) are the peak electric field amplitude and angular frequency of the driving laser pulse. 

\paragraph{Driving electric field.}  
The atom is driven by a laser pulse consisting of a noiseless fundamental component at frequency \(\omega\) ($\lambda \approx$ 800 nm) and a stochastic component at half the frequency (\(\omega/2\)):  

\[
E(t) = \Big[ y_x \cos\!\Big(\tfrac{\omega t}{2} + \phi\Big) + y_y \sin\!\Big(\tfrac{\omega t}{2} + \phi\Big) + E_0 \cos(\omega t) \Big] \, f(t).
\]

In which \(E_0\) corresponds to a peak intensity of \(I = 10^{14}\,\mathrm{W/cm^2}\) in atomic units, \(f(t)\) is the pulse envelope (defined below), and \(\phi\) is the relative phase between the \(\omega\) and \(\omega/2\) components (i.e., the two-color phase), scanned over \([0,2\pi]\), and \(y_x\) and \(y_y\) are the electric field amplitudes of the cosine and sine quadratures of stochastic \(\omega/2\) component; 

\paragraph{Pulse envelope implementation.}  
A trapezoid envelope \(f(t)\) is constructed by concatenating a time-linear rising segment, a flat plateau, and a linear fall. With $N_{\mathrm{rise}}$ rise cycles, $N_p$ plateau cycles, and $N_{\mathrm{fall}}$ fall cycles, the total pulse duration is  

\[
T = (N_{\mathrm{rise}}+N_p+N_{\mathrm{fall}})\,T_{\mathrm{cyc}}, \qquad T_{\mathrm{cyc}} = \frac{2\pi}{\omega}.
\]
with \(N_{rise}=N_{fall}=3, N_{p}=8\). Defining $t_0 = 0$, $t_{\mathrm{rise}} = t_0 + N_{\mathrm{rise}}T_{\mathrm{cyc}}$, and $t_{\mathrm{fall}} = T - N_{\mathrm{fall}}T_{\mathrm{cyc}}$, the envelope is  

\[
f(t) = 
\begin{cases} 
\frac{t-t_0}{t_{\mathrm{rise}}-t_0}, & t_0 < t < t_{\mathrm{rise}}, \\[6pt]
1, & t_{\mathrm{rise}} < t < t_{\mathrm{fall}}, \\[6pt]
\frac{T-t}{T-t_{\mathrm{fall}}}, & t_{\mathrm{fall}} < t < T, \\[6pt]
0, & \text{otherwise}.
\end{cases}
\]  

This provides a trapezoidal envelope that smoothly switches the field on and off.  

\paragraph{Stochastic sampling of the $\omega/2$ field.}  
We sample the amplitude of the $\omega/2$ quadratures from independent Gaussian distributions:  

\[
y_x \sim \mathcal{N}(0, \sigma_x^2), \qquad 
y_y \sim \mathcal{N}(0, \sigma_y^2),
\]  
With the choice:

\[
\sigma_x = \frac{2.5\times 10^{5}\,\varepsilon^{(1)}}{\sqrt{2}}, 
\qquad \sigma_y = \frac{2.5\times 10^{4}\,\varepsilon^{(1)}}{\sqrt{2}}, 
\qquad \varepsilon^{(1)} \approx 10^{-8}.
\]  

This mimics vacuum (for \(\sigma_1=\sigma_2=\epsilon^{(1)}\)) or bright squeezed-vacuum statistics in the \(\omega/2\) mode. Each pair \((y_x, y_y)\) defines a random quadrature realization of the low-frequency field.  

\paragraph{Propagation.}  
The wavefunction is propagated with a split-step operator method, from which we obtain the time dependent dipole acceleration as
\[
a(t) = \big\langle -\partial_x V_{\mathrm{atom}}(x) \big\rangle - E(t).
\]

The Fourier transform \(a(\omega)\) provides the single-atom harmonic emission spectrum for each shot, that is in good qualitative agreement with the experimental results, up to an overall scale factor; the overall scale factor may be attributed to the choice of the amplitude of infrared vacuum fluctuations in the simulation, \(\epsilon^{(1)}\), which is not measured experimentally. 


\subsubsection*{2. Reconstruction of quadrature distributions, Husimi, and Wigner functions}  

\paragraph{Quadrature extraction.}  
For a given harmonic order \(H\), we extract the Fourier coefficient from the acceleration trace \(a(H\omega)\):  
\[
a(H\omega) \propto X + i P,
\] 

where  
\[
X = \Re(a(H\omega)), \qquad P = \Im(a(H\omega)).
\]  
Each Monte-Carlo shot contributes a pair \((X,P)\), forming an ensemble of quadrature samples.  

\paragraph{Normalization to vacuum.}  
To assign physical meaning to the quadratures, we normalize them against a reference dataset, calculated with \(\sigma_1=\sigma_2=\epsilon^{(1)}\). For each harmonic we compute 
\[
\sigma_X^2 = \mathrm{Var}(X), \qquad \sigma_P^2 = \mathrm{Var}(P),
\] and rescale (X,P) such that for vacuum, 

\[
\frac{1}{2}\big(\sigma_X^2 + \sigma_P^2\big) = \tfrac{1}{2}.
\]
\paragraph{Husimi Q function.}  
The normalized quadratures are binned into a 2D histogram, interpolated onto a fine grid, and convolved with a Gaussian kernel for smoothing. This yields the Husimi Q distribution:  

\[
Q_H(X,P) = \frac{1}{\pi} \langle \alpha | \hat{\rho}_H | \alpha \rangle, 
\qquad \alpha = \frac{X+iP}{\sqrt{2}}.
\]


	\begin{figure}
		\centering
        \includegraphics[width=0.95\linewidth]{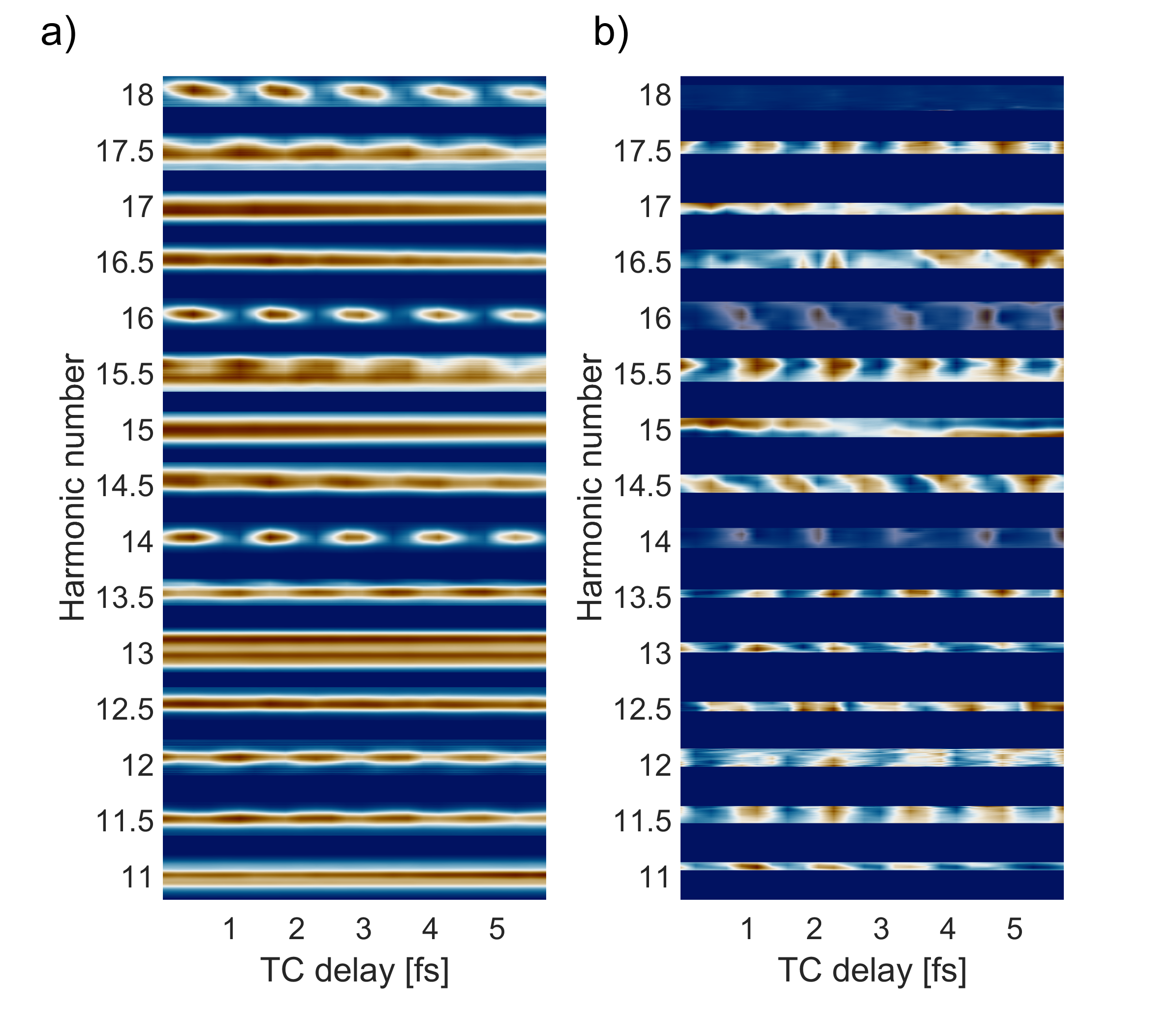 }
		\caption{ {\bf Oscillations of the statistics of the attosecond light as a function of the two color delay.} \bf{a)} intensity mean value and \bf{b} \gt as a function of the two-color delay, normalized separately for each harmonic, and DC normalized with respect to the delay.	}
			\label{dynamicFigure}
	\end{figure}

\begin{figure}[!htb]
  \centering
  \includegraphics[width=0.95\columnwidth]{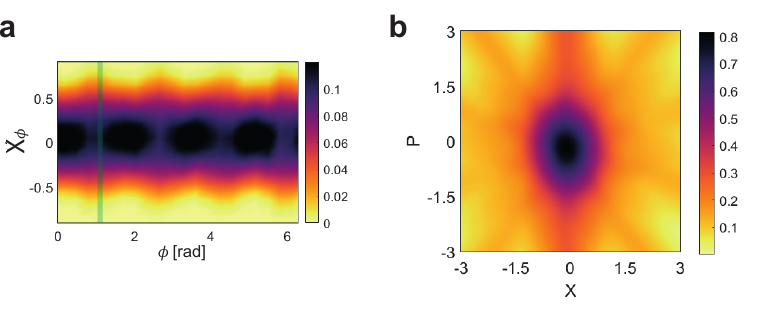}
  \caption{\textbf{Tomography of the Wigner function single-photon channel for a half-integer harmonic.}
  \textbf{(a)} Measured quadrature amplitude $X_\phi$ of harmonic $q=14.5$ versus two-color delay $\phi_{\mathrm{TC}}$. The mean displacement is negligible and the variance oscillates with $\phi_{\mathrm{TC}}$, consistent with squeezed-like behavior.
  \textbf{(b)} Reconstructed Wigner function obtained from the set $\{P(X_\phi)\}_\phi$ via inverse Radon transform. The distribution is centered near zero displacement and exhibits anisotropic noise, characteristic of squeezed-vacuum. }
  \label{fig:wigner_single_photon}
\end{figure}


\clearpage 



\end{document}